\documentclass[reprint,amsmath,amssymb,amsbsy,prb,reprintnumbers,showpacs]{revtex4-2}
\usepackage{graphicx,color}
\usepackage{dcolumn}
\usepackage{bm}
\usepackage{braket}
\usepackage{mathtools}
\usepackage{ulem}
\usepackage[breaklinks,colorlinks=true,linkcolor=blue,urlcolor=blue,citecolor=blue]{hyperref}
\usepackage{times}
\usepackage{filecontents}

\begin{document}
\title{Molecular-orbital representation with random U(1) variables}
\author{Tomonari Mizoguchi}
\affiliation{Department of Physics, University of Tsukuba, Tsukuba, Ibaraki 305-8571, Japan}
\email{mizoguchi@rhodia.ph.tsukuba.ac.jp}
\author{Yasuhiro Hatsugai}
\affiliation{Department of Physics, University of Tsukuba, Tsukuba, Ibaraki 305-8571, Japan}
\date{\today}

\begin{abstract}
We propose random tight-binding models that host macroscopically degenerate zero energy modes and belong to the unitary class.
Specifically, we employ the molecular-orbital representation, 
where a Hamiltonian is constructed by a set of non-orthogonal orbitals composed of linear combinations of atomic orbitals. 
By setting the coefficients appearing in molecular orbitals to be random U(1) variables, 
we can make the models belong to the unitary class.
We find two characteristic behaviors that are distinct from the random-real-valued molecular-orbital model. 
Firstly, a finite energy gap opens on top of the degenerate zero energy modes.
Secondly, besides the zero energy modes, we also argue that the band center of the finite energy modes is critical, 
which is inherited from the dual counterpart, namely, the random-phase model on a bipartite lattice. 
Furthermore, as a by-product of this model-construction scheme, 
we also construct the random tight-binding model 
on a composite lattice, 
where we also find a realization of critical states. 
\end{abstract}

\maketitle
\section{Introduction}
The effects of disorders on electronic systems have been a central issue in condensed matter physics. 
Metal-insulator transition induced by disorders, i.e., the Anderson localization, is one of the most striking 
phenomena caused by disorders~\cite{Anderson1958,Abrahams1979}, and 
there have been tremendous amount of research activities since its proposal. 
More recently, the effects of disorders on systems with exotic electronic structures have been studied extensively. 
For instance, Dirac fermions, characterized by vanishing density of states at the Dirac points,
exhibit rich physics in the presence of disorders~\cite{Hatsugai1993,Chamon1996,Hatsugai1997,Castillo1997,Morita1997,Fukui2003}.
Another example of exotic electronic structures
 is flat bands, i.e., the completely dispersionless bands in the entire Brillouin zone.
In flat-band systems, macroscopic degeneracy of the single-particle spectrum allows us to construct localized eigenstates even without disorders~\cite{Zhitomirsky2004,Bergman2008,Kuno2020}.
Then, they are expected to be highly sensitive to disorders. 
Indeed, various characteristic phenomena due to the interplay between flat bands and disorders have been reported~\cite{Goda2006,Nishino2007,Chalker2010,Leykam2013,Shukla2018,Shukla2018_2,Bilitewski2018,Kuno2021}. 

Besides the electronic structures, the symmetry and dimensionality 
also play crucial a role in determining the nature of the Anderson localization~\cite{Altland1997}.
In theoretical analysis of tight-binding models, the type of disorders 
(e.g., random potentials, random hoppings, random flux, etc.) 
and the lattice structures are two key ingredients for determination of the symmetry class.
In this respect, the previous works on disordered flat-band systems mainly focus on the
case where random on-site potentials are introduced to a flat band model,
and a Hamiltonian matrix is consequently a real matrix~\cite{Goda2006,Nishino2007,Chalker2010,Kuno2021}. 
In contrast to those works, we have developed yet another direction of studying random flat-band models, 
that is, tailoring the models such that macroscopic degeneracy can be exactly retained even in the presence of disorders.
To be more specific, we have studied tight-binding Hamiltonians
that are written down by a set of non-orthogonal and unnormalized orbitals constructed by a linear combination of 
atomic orbitals (AOs) and whose total number is smaller than that of the AOs~\cite{remark}. 
We name such construction the ``molecular-orbital" (MO) representation~\cite{Hatsugai2011,Hatsugai2015,Mizoguchi2019,Mizoguchi2020,Mizoguchi2021_skagome,Hatsugai2021,Mizoguchi2022,Kuroda2022}.
Along this line, in the previous works, 
we consider the MO models where the coefficients appearing in MOs 
are random real values~\cite{remark_mo}.
We refer to this type of models as the real-valued random MO models henceforth.
We have found that the probability density of zero energy modes is seemingly characteristic and resemble that of the critical state,
but its scaling behavior upon changing the system size obeys that of the extended state~\cite{Hatsugai2021,Kuroda2022}. 

\begin{figure}[b]
\begin{center}
\includegraphics[clip,width = 0.98\linewidth]{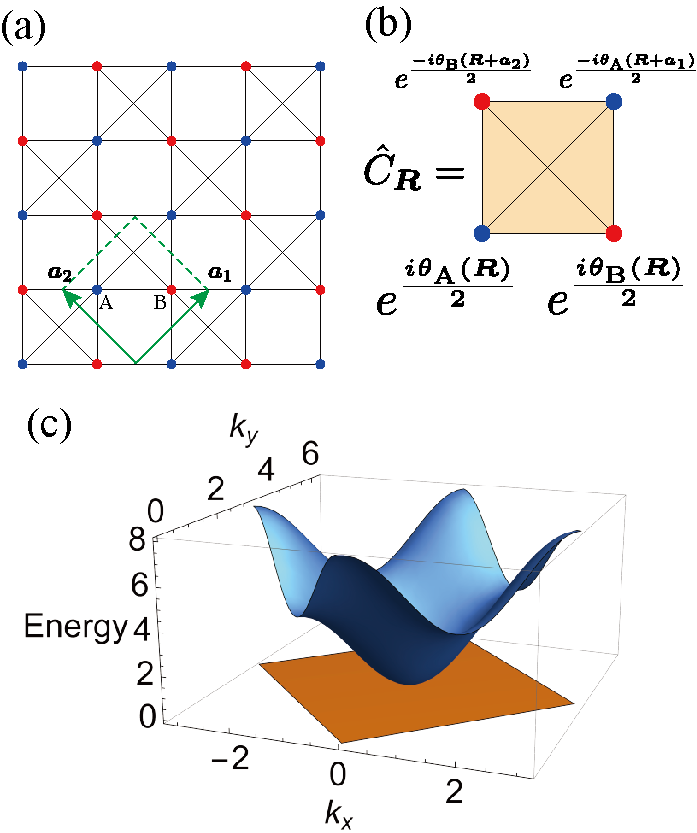}
\vspace{-10pt}
\caption{
(a) Checkerboard lattice. Two lattice vectors are $\bm{a}_1 = (1,1)$ and $\bm{a}_2 = (-1,1)$.
(b) Schematic figure of $\hat{C}_{\bm{R}}$. 
(c) The dispersion relation for the clean limit (i.e., $\theta_{\rm A/B} (\bm{R})= 0$).}
  \label{fig:ch_clean}
 \end{center}
 \vspace{-10pt}
\end{figure}

With this as the background, in this paper, we consider the random MO models with different type of 
randomness from that in the previous works.
Specifically, we construct the MOs such that the coefficients are random U(1) variables, 
which breaks the time-reversal symmetry and thus the universality class becomes unitary. 
We refer to the models constructed in this way as the random-phase MO models. 

We first study the checkerboard model to illustrate the idea of model construction. 
A key feature is that this model is 
a dual of the random-phase (also referred to as the random vector potential or the random-flux) square lattice model,
which has been studied intensively as a prime example of the random lattice fermion models in the unitary class
~\cite{Lee1981,Pryor1992,Kalmeyer1993,Sugiyama1993,Ohtsuki1993,Avishai1993,Gade1993,Ludwig1994,Miller1996,Furusaki1999}.

By the numerical calculations, we elucidate the similarities and differences between the random-phase MO models 
and the real-valued random MO models.
On the energy spectrum, a finite energy gap opens on top of the degenerate zero-energy modes, 
which is in contrast to the real-valued random MO model.
Meanwhile, the probability density distribution of the zero-energy 
modes has similar behavior to that of the real-valued random MO model in that
its finite size scaling behavior of the inverse participation ratio (IPR) obeys that of extended states.
Besides the zero-energy modes, we also find a characteristic feature in the finite-energy modes,
namely, the band center of the finite-energy modes is a critical state, which is inherited from the random-phase square lattice model.
Notably, the random-phase MO model itself is not chiral symmetric, 
although the chiral symmetry plays an important role in the emergence of the critical state in the random-phase square lattice model.

We additionally show that the random-phase MOs 
used to describe the checkerboard model can be used 
to construct a model defined on a composite lattice of the square and checkerboard lattices, i.e., the Lieb lattice.
In fact, the authors have applied a similar model construction
scheme to a kind of topological insulators, called the square-root topological insulators~\cite{Mizoguchi2020_sq,*Mizoguchi_erratum,Mizoguchi2021}.
We show that, in the Lieb-lattice model, the critical states inherited 
from the square lattice random-phase model appear at the center of the positive and negative energy sectors in a pairwise manner.  

Subsequently to the checkerboard and Lieb-lattice models, 
we also apply this construction method to the kagome model, 
where we find that the results are qualitatively the same as those of the checkerboard model.
In addition, we also argue the composite-lattice model of this series;
the corresponding composite lattice model is a decorated honeycomb lattice model,
where we again find the qualitatively same results as those of the Lieb-lattice model. 
These results indicate the ubiquity of the present model construction by the random-phase MO representation. 

The rest of this paper is organized as follows.
In Sec.~\ref{sec:checker}, we introduce the random-phase MO model on a checkerboard lattice,
and elucidate its basic properties, including the dual relation to the random-phase square-lattice model.
Then, in Sec.~\ref{sec:numerical}, we present our numerical results on the spectrum and wave functions.
We focus on two aspects, namely, the gap opening between the degenerate zero-energy modes and the lowest finite energy modes, and 
the critical nature of the center of the finite energy modes.
In Sec.~\ref{sec:lieb}, we argue that the random-phase MO introduced to describe the checkerboard model can be used to
construct 
the composite-lattice model, 
namely, the Lieb-lattice model.
In Sec.~\ref{sec:kagome}, we apply the same construction to the kagome lattice, and 
we show the parallel numerical results to the checkerboard model. 
The results for the composite lattice, namely, the decorated honeycomb model, are also shown. 
The summary of this paper is presented in Sec.~\ref{sec:summary}.

\section{Model construction scheme: Checkerboard lattice \label{sec:checker}}
In this section, we introduce a random-phase MO model on a checkerboard lattice [Fig.~\ref{fig:ch_clean}(a)].
The system consists of $L \times L (=N_{\rm u.c.})$ unit cells and the periodic boundary condition is imposed. 
At each site $i$ of the checkerboard lattice, we define the AO whose annihilation operator is written as $c_i$.
The site $i$ is specified by the unit-cell position $\bm{R} = R_1 \bm{a}_1 + R_2 \bm{a}_2$ and the sublattice index A/B. 

On this lattice, we consider a tight-binding Hamiltonian:
\begin{eqnarray}
H = \sum_{\bm{R}} \hat{C}^\dagger_{\bm{R}}  \hat{C}_{\bm{R}},  \label{eq:ham_checker}
\end{eqnarray}
where 
\begin{eqnarray}
 \hat{C}_{\bm{R}} &=& e^{\frac{i \theta_{\rm A}(\bm{R})}{2}} c_{\bm{R},\mathrm{A}} 
 + e^{\frac{i \theta_{\rm B}(\bm{R})}{2}} c_{\bm{R},\mathrm{B}}  \nonumber \\
 &+& e^{- \frac{i \theta_{\rm A}(\bm{R} + \bm{a}_1)}{2}} c_{\bm{R}
 +\bm{a}_1,\mathrm{A}} + e^{- \frac{i \theta_{\rm B}(\bm{R} + \bm{a}_2)}{2}} c_{\bm{R}+\bm{a}_2,\mathrm{B}}. \label{eq:MO_choice}
\end{eqnarray}
See Fig.~\ref{fig:ch_clean}(b) for the schematic figure of $\hat{C}_{\bm{R}}$.
It is worth noting that 
the MOs are defined on lattice sites of a square lattice which are placed at the center of the crossed squares of a checkerboard lattice. 
In Eq.~(\ref{eq:MO_choice}), 
the phase $\theta_{\rm A/B}(\bm{R}) 
\in [-\pi, \pi]$ is a random variable obeying the uniform distribution. 
The choice of MOs in Eq.~(\ref{eq:MO_choice}) might looks fine-tuned because the phase factors appearing in
neighboring MOs are not independent. 
For instance, the variable $\theta_{\rm A}(\bm{R})$ appears both in $\hat{C}_{\bm{R}}$ and in $\hat{C}_{\bm{R}-\bm{a}_1}$.
In Appendix~\ref{app:remark}, we elucidate the relation between this model and the model where all phase factors are random variables.
We call $\hat{C}_{\bm{R}}$ the MO since it consists of local linear combination of AOs. 

We align all the AOs in a column vector form, which we write as $\hat{\bm{c}}$;
similarly, we align the MO in a column vector form, which we write as $\hat{\bm{C}}$.
Using these, 
we can rewrite the Hamiltonian of Eq.~(\ref{eq:ham_checker}) as 
\begin{eqnarray}
 H = \hat{\bm{c}}^\dagger \mathcal{H}  \hat{\bm{c}}, \mathcal{H} = \Psi \Psi^\dagger, \label{eq:Ham_MO_CH}
\end{eqnarray}
where $\Psi$ is a $2 N_{\rm u.c.} \times N_{\rm u.c.}$ matrix 
defined such that it satisfies $\hat{\bm{C}}^\dagger = \hat{\bm{c}}^\dagger\Psi$. 
Then, the single-particle eigenenergies and eigenstates are obtained by solving the eigenvalue equation of the matrix $\mathcal{H}$.

In fact, Eq.~(\ref{eq:Ham_MO_CH}) indicates that there are at least $N_{\rm u.c.}$ zero-energy eigenstates of $\mathcal{H}$.
To be specific, as $\Psi^\dagger$ is the non-square matrix where the number of 
rows is lesser than that of columns, the dimension of the kernel of the linear map expressed by $\Psi^\dagger$ 
is equal to or greater than $N_{\rm u.c.}$.
In other words, there exist the vectors $\bm{\phi}^{\rm ZM}_\ell$ ($\ell = 1, \cdots, N_{\rm u.c.}$) 
such that 
$\bm{\phi}^{\rm ZM}_\ell$ satisfies $\Psi^\dagger \bm{\phi}^{\rm ZM}_\ell= 0$. 
Then, such vectors also satisfy $\mathcal{H} \bm{\phi}^{\rm ZM}_\ell =  \Psi (\Psi^\dagger \bm{\phi}^{\rm ZM}_\ell) = 0$.
In the following, we assume that $\bm{\phi}^{\rm ZM}_\ell$ is normalized, namely,
$(\bm{\phi}^{\mathrm{ZM}}_\ell)^\dagger \bm{\phi}^{\rm ZM}_\ell =1$ holds. 
\begin{figure}[tb]
\begin{center}
\includegraphics[clip,width = 0.8\linewidth]{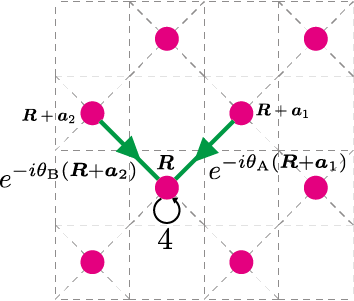}
\vspace{-10pt}
\caption{
Schematic figure of the square-lattice tight-binding model with random phase factor, 
corresponding to $\Upsilon$.}
  \label{fig:dual}
 \end{center}
 \vspace{-10pt}
\end{figure}

\subsection{Review of clean-limit properties \label{sec:clean}}
Without phase factors, i.e., $\theta_{\rm A}(\bm{R}) = \theta_{\rm B}(\bm{R}) = 0$, 
the model is the conventional checkerboard lattice model with the nearest-neighbor hoppings being $1$ and the on-site potential being $2$. 
In this case, the eigenenergy of the dispersive band as a function of a crystal momentum
is easily obtained as $E_{\bm{k}} = 4 + 2 (\cos \bm{k}\cdot \bm{a}_1 + \cos \bm{k}\cdot \bm{a}_2)$;
the other band is the flat band with the eigenenergy being 0.
The band structure is depicted Fig.~\ref{fig:ch_clean}(c), where we see the quadratic band touching
at $\bm{k} =\frac{\bm{b}_1 + \bm{b}_2}{2} = (0,\pi)$ [$\bm{b}_1 = (\pi, \pi)$ and $\bm{b}_2 = (-\pi, \pi)$ are the reciprocal lattice vectors].
Recently, several theoretical understandings of this type of band touching were proposed~\cite{Bergman2008,Bilitewski2018,Rhim2019,Hwang2021,Hwang2021_2,Graf2021}. 
From the viewpoint of the MO representation, 
the origin of the quadratic band touching is the linear dependence of the MOs. 
To be specific, in the clean limit, the following relation holds:
\begin{eqnarray}
\sum_{\bm{R}} (-1)^{R_1 + R_2} \hat{C}_{\bm{R}} = 0. \label{eq:MO_linear}
\end{eqnarray}
This means that the number of linearly independent MOs (or the rank of $\Psi^\dagger$) is $N_{\rm u.c.}-1$, rather than $N_{\rm u.c.}$.
Recalling the fact that the zero modes belong to the kernel of $\Psi^\dagger$, 
the degeneracy of zero modes is $N_{\rm u.c.}+1$, which leads to band touching of the bottom of the dispersive band
to the flat band. 
\begin{figure*}[tb]
\begin{center}
\includegraphics[clip,width = 0.95\linewidth]{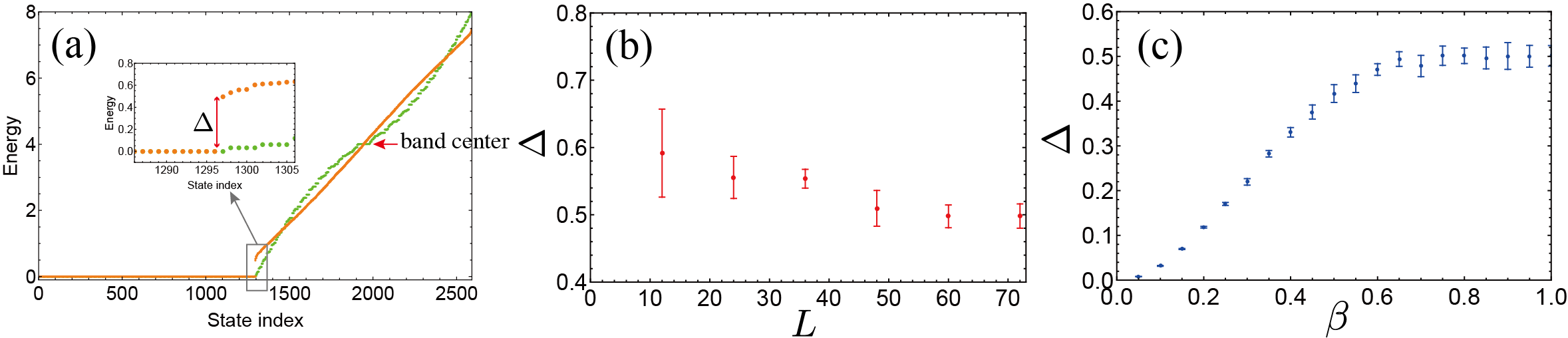}
\vspace{-10pt}
\caption{
(a) Energy spectrum for the system with $L = 36$. 
Orange (green) dots correspond to the random (clean) system.
(b) The gap $\Delta$ as a function of $L$. 
(c) The gap as a function of $\beta$ for $L=60$. 
For (b) and (c), the average and the standard deviation are calculated for ten independent configurations of $\theta$'s. }
  \label{fig:ch_en}
 \end{center}
 \vspace{-10pt}
\end{figure*}

\subsection{Semi-positivity of $\mathcal{H}$}
\begin{figure*}[tb]
\begin{center}
\includegraphics[clip,width = 0.95\linewidth]{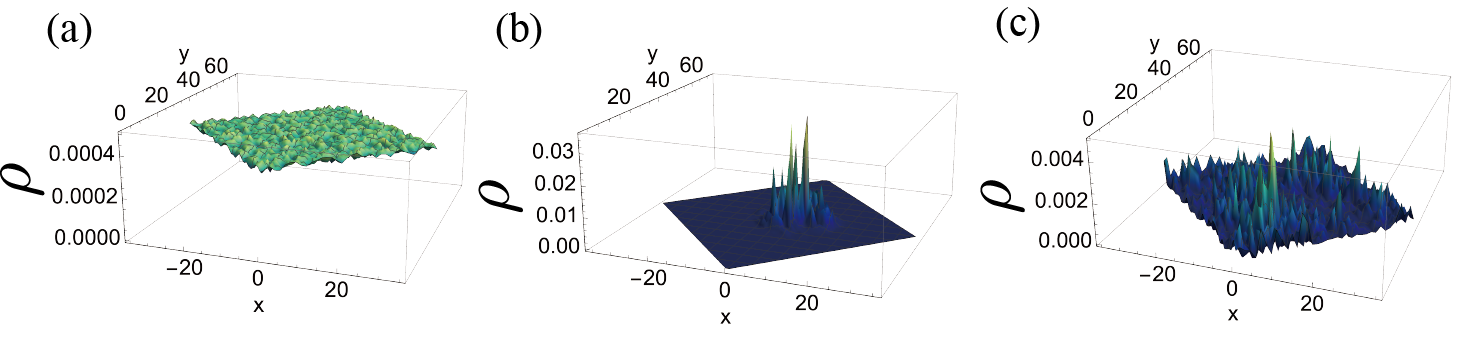}
\vspace{-10pt}
\caption{The probability density for (a) the zero modes, (b) the lowest finite-energy mode,
and (c) the center of the finite-energy modes.
We set $L= 36$. 
}
  \label{fig:ch_probdens}
 \end{center}
 \vspace{-10pt}
\end{figure*}
We now turn to the disordered case. 
Firstly, it is important to point out that, from Eq.~(\ref{eq:Ham_MO_CH}), 
$\mathcal{H}$ is positive semi-definite for generic random phases.
A straightforward proof of this is as follows: let $\bm{\phi}$ be an arbitrary $2N_{\rm u.c.}$-component 
column vector. 
Then, we have $\bm{\phi}^\dagger \mathcal{H} \bm{\phi} = |\Psi^\dagger \bm{\phi}|^2 \geq 0$,
which indicates the semi-positivity of $\mathcal{H}$.
This is enough, but here we present an alternative proof, which gives a useful insight on $\Psi$ and $\Psi^\dagger$
(see Sec.~\ref{sec:dual_relation} for details). 
We introduce a Hermitian matrix
\begin{eqnarray}
\bar{\mathcal{H}} = 
\begin{pmatrix}
\mathcal{O}_{N_{\rm u.c.},N_{\rm u.c.} } & \Psi^\dagger \\
\Psi & \mathcal{O}_{2 N_{\rm u.c.},2 N_{\rm u.c.}}\\
\end{pmatrix}, \label{eq:Ham_Lieb}
\end{eqnarray}
where $\mathcal{O}_{n,m}$ denotes a $n\times m$ zero matrix. 
Then, the square of the matrix $\bar{\mathcal{H}} $ is block-diagonalized as 
\begin{eqnarray}
\bar{\mathcal{H}}^2  = 
\begin{pmatrix}
 \Upsilon & \mathcal{O}_{N_{\rm u.c.},2 N_{\rm u.c.}} \\
\mathcal{O}_{2 N_{\rm u.c.}, N_{\rm u.c.}} &  \mathcal{H} \\
\end{pmatrix}, \label{eq:Hsq}
\end{eqnarray}
where $ \Upsilon = \Psi^\dagger \Psi$. 
The matrix $\bar{\mathcal{H}}^2$ is obviously positive semi-definite, 
and so are $\Upsilon$ and $\mathcal{H}$, 
since they are top-left and bottom-right blocks of $\bar{\mathcal{H}}^2$, respectively.

\subsection{Relation to the square-lattice model and its implication \label{sec:dual_relation}}
In the following discussions, $\Upsilon$ of Eq.~(\ref{eq:Hsq}) plays a crucial role in 
understanding the properties of $\mathcal{H}$.
In fact, the eigenvalues of $\Upsilon$ are in common with those of $\mathcal{H}$.
This is simply understood by the following relation~\cite{DiFrancesco1990,Pearce1993,Mizoguchi2021_Diamond}:
\begin{eqnarray}
\mathcal{H} \Psi = \Psi \Upsilon.
\end{eqnarray} 
Using this relation, we find the following: Let $\bm{u}_\ell$ 
be a normalized eigenvector of $\Upsilon$
with the eigenvalue $\varepsilon_\ell$.
Then, we have 
\begin{eqnarray}
\mathcal{H} (\Psi \bm{u}_\ell) = \Psi ( \Upsilon \bm{u}_\ell ) = \varepsilon_\ell (\Psi \bm{u}_\ell ),
\end{eqnarray}
meaning that $\Psi \bm{u}_\ell$ is an eigenvector of $\mathcal{H}$ with the eigenvalue $\varepsilon_\ell$,
up to the normalization constant.
Further, noting that
$|\Psi \bm{u}_\ell |^2 = \bm{u}_\ell^\dagger \Psi^\dagger \Psi \bm{u}_\ell = \bm{u}_\ell^\dagger \Upsilon \bm{u}_\ell = \varepsilon_\ell$,
we find that the normalized eigenvector of $\mathcal{H}$ with the eigenenergy $\varepsilon_\ell$ is
\begin{eqnarray}
\bm{\phi}_{\ell}^{\rm NZM}  = \frac{1}{\sqrt{\varepsilon_\ell}} \Psi \bm{u}_\ell. 
\label{eq:eigen_h}
\end{eqnarray}

Importantly, in the present model, $\Upsilon$ is given as 
\begin{eqnarray}
\Upsilon= 4\hat{I}_{N_{\rm u. c.}} + \tilde{\mathcal{H}}^{\rm sq}, \label{eq:Up_sq}
\end{eqnarray}
where $\hat{I}_{n}$ represents the $n \times n$ identity matrix and 
$\tilde{\mathcal{H}}^{\rm sq}$ corresponds to the square-lattice tight-binding model with the random phase factor (Fig.~\ref{fig:dual}): 
\begin{eqnarray}
[\tilde{\mathcal{H}}^{\rm sq}]_{\bm{R},\bm{R}^\prime} =  
\left\{
\begin{array}{ll}
e^{-i\theta_{\rm A}(\bm{R}+\bm{a}_1)}, & \bm{R}^\prime = \bm{R}+\bm{a}_1  \\
e^{i\theta_{\rm A}(\bm{R})}, &  \bm{R}^\prime = \bm{R}-\bm{a}_1 \\
e^{-i\theta_{\rm B}(\bm{R}+\bm{a}_2)}, & \bm{R}^\prime = \bm{R}+\bm{a}_2  \\
e^{i\theta_{\rm B}(\bm{R})}, &  \bm{R}^\prime = \bm{R}-\bm{a}_2 \\
0, & \mathrm{otherwise} \\
\end{array}
\right.
.
\end{eqnarray}
This model of the square-lattice random phase (or the random flux) model, has been intensively studied in the literature~\cite{Lee1981,Pryor1992,Kalmeyer1993,Sugiyama1993,Ohtsuki1993,Avishai1993,Gade1993,Ludwig1994,Miller1996,Furusaki1999}.
Note that $\tilde{\mathcal{H}}^{\rm sq}$ is chiral symmetric, that is, 
$\tilde{\mathcal{H}}^{\rm sq}$ 
satisfies $\{\tilde{\mathcal{H}}^{\rm sq}, \Gamma \} =0$
with $[\Gamma]_{\bm{R},\bm{R}^\prime}  = (-1)^{R_1 + R_2 } \delta_{\bm{R},\bm{R}^\prime}$~\cite{remark_chiral}.

We show that the semi-positivity of $\Upsilon$ we addressed in the previous subsection leads to an important consequence. 
To be specific, as $\Upsilon$ is positive semi-definite, the smallest eigenvalue of $\tilde{\mathcal{H}}^{\rm sq}$,
$\varepsilon^{\rm sq}_{\rm min}$, satisfies $\varepsilon^{\rm sq}_{\rm min} \geq -4$. 
In addition, as $\tilde{\mathcal{H}}^{\rm sq}$ is chiral symmetric, 
its largest eigenvalue is $\varepsilon^{\rm sq}_{\rm max}$ is -$\varepsilon^{\rm sq}_{\rm min}$,
so it satisfies $\varepsilon^{\rm sq}_{\rm max} \geq 4$. 
As $-4$ and $4$ are the minimum and maximum of the eigenvalues for the square-lattice tight-binding model without
the phase factor, 
the above facts provide a proof that the expansion of the band width by the introduction of the phase factor is prohibited. 

\section{Numerical results for the checkerboard model \label{sec:numerical}}
In this section, we present our numerical results for the random phase 
MO model on a checkerboard lattice of Eq.~(\ref{eq:Ham_MO_CH}). 
\subsection{Energy spectrum}
\begin{figure}[tb]
\begin{center}
\includegraphics[clip,width = 0.95\linewidth]{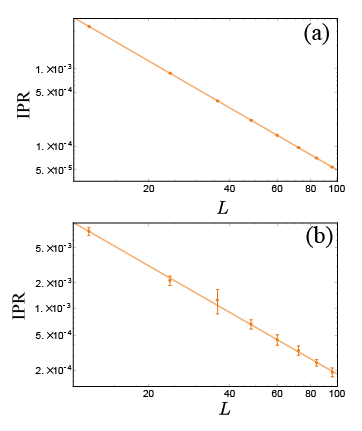}
\vspace{-10pt}
\caption{IPR for 
(a) the degenerate zero modes
and 
(b) the center of the finite-energy modes.
The line represent the fitting function, IPR$=A\cdot L^{-B}$,
with (a) $A=0.50$, $B=1.99$
and (b) $A=0.60$, $B=1.76$. 
The average and the standard deviation are calculated for ten independent configurations of $\theta$'s.}
  \label{fig:ch_ipr}
 \end{center}
 \vspace{-10pt}
\end{figure}
In Fig.~\ref{fig:ch_en}(a), we show an energy spectrum in an ascending order for $L=36$. 
The orange dots correspond to the random phase case, while the green dots correspond to the clean limit. 
We clearly see degenerate zero-energy modes, whose degeneracy is $N_{\rm u.c.} = L^2$. 
On top of the zero-energy modes, there exists a finite energy gap, which is in sharp contrast to the clean limit,
and to the real-valued random MO model~\cite{Hatsugai2021,Kuroda2022}.
In Fig.~\ref{fig:ch_en}(b), we show the gap as a function of $L$, which indicates that the energy gap 
does not vanish when extrapolated to $L \rightarrow \infty$. The gap size is about $0.5$. 
Recalling the relation between $\Upsilon$ and $\mathcal{H}$ and the relation of Eq.~(\ref{eq:Up_sq}),
we find that the gap $\Delta$ corresponds to the ``shrinking" of the band width  due to the random phase in the square-lattice tight-binding model.
This shrinking of the band width was observed in the numerical simulations in the literature~\cite{Pryor1992,Ohtsuki1993,Sugiyama1993,Furusaki1999}.

As we have seen in Sec.~\ref{sec:clean}, the gap remains closed in the clean limit,
so one may wonder how the gap evolves upon increasing the strength of disorders. 
To see this, we interpolate the disordered limit and the clean limit, 
To be specific, we modify the range of $\theta$ as 
$\theta_{\rm A/B}(\bm{R}) \in [-\beta \pi, \beta \pi]$
with $\beta \in [0,1]$.
Figure.~\ref{fig:ch_en}(c) represents the $\beta$ dependence of $\Delta$ for $L = 60$.
We see that $\Delta$ is smoothly dependent on $\beta$, and it saturates around $\beta \sim 0.6$.

\subsection{Wave functions: Emergent critical state at band center}
\begin{figure*}[tb]
\begin{center}
\includegraphics[clip,width = 0.95\linewidth]{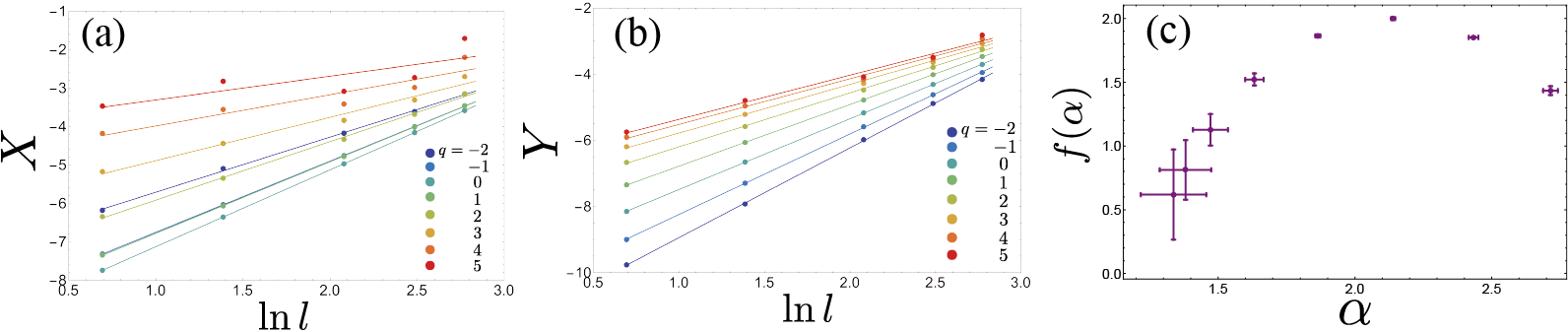}
\vspace{-10pt}
\caption{
Multifractal analysis for the center of the finite-energy modes.
(a) $X$ [Eq.~(\ref{eq:X})] and (b) $Y$ [Eq.~(\ref{eq:Y})] as functions of $\ln l$ for several values of $q$.
(c) $f$-$\alpha$ spectrum.
The error bars in the panel (c) are the fitting errors 
for $X$ and $Y$.}
  \label{fig:ch_fa}
 \end{center}
 \vspace{-10pt}
\end{figure*}
Next, we reveal the characteristics of the wave functions for both degenerate zero modes and the finite energy modes. 
Figure~\ref{fig:ch_probdens} shows the probability density, $\rho$. 
The definition of the probability density is as follows.
For the finite energy mode, it is defined as 
\begin{eqnarray}
\rho^{\rm NZM}_{\ell} (i) = |[\bm{\phi}^{\rm NZM}_{\ell} ]_i |^2, \label{eq:PD_NZM}
\end{eqnarray}
where $\bm{\phi}^{\rm NZM}_{\ell}$ ($\ell = 1, \cdots, N_{\rm u.c.}$) is the $\ell$-th 
non-zero energy eigenvector of $\mathcal{H}$ [see Eq.~(\ref{eq:eigen_h})]; 
here we align them such that they satisfy 
$\varepsilon_1 \leq \varepsilon_2 \leq \cdots \leq \varepsilon_{N_{\rm u.c.}}$. 
For the degenerate zero-energy modes, $\rho$ is defined as,
\begin{eqnarray}
\rho^{\rm ZM}(i) =\frac{1}{N_{\rm u.c.}} \sum_{\ell = 1}^{N_{\rm u.c.}} 
|[\bm{\phi}^{\rm ZM}_{\ell} ]_i |^2.
\end{eqnarray}
Note that the probability density satisfies $\sum_{i} \rho^{\rm NZM}_{\ell} (i)  = \sum_i \rho^{\rm ZM}(i) = 1$,
since the wave functions are normalized.
We also note that $\rho^{\rm NZM}_{\ell} (i)$ is related to the correlation matrix defined for $\bm{u}_{\ell}$
(see Appendix~\ref{app:pd_cm} for details).

Figures~\ref{fig:ch_probdens}(a), \ref{fig:ch_probdens}(b), and \ref{fig:ch_probdens}(c) represent $\rho$
for the zero modes, the non-zero mode of $\ell =1$ (which we call the band edge), and the non-zero mode of $\ell =\frac{N_{\rm u.c.}}{2}$ (which we call the band center),
respectively. 
For the zero modes, we see that the distribution is rather uniform despite the existence of the randomness.
At the band edge ($\ell =1$), we see a sharp peak of the probability density, 
indicating the localized nature of the wave function.
Interestingly, the at the band center ($\ell =\frac{N_{\rm u.c.}}{2}$), we see a spiky distribution of the probability density,
indicating the critical nature of the wave function.
In fact, the possibility of the critical state at the band center is inferred from
from Eq.~(\ref{eq:eigen_h}).
To be more specific, $\bm{u}_{\frac{N_{\rm u.c.}}{2}}$, 
which is the wave function of the band center of $\tilde{\mathcal{H}}^{\rm sq}$,
is predicted to be critical~\cite{Lee1981,Pryor1992,Kalmeyer1993,Sugiyama1993,Ohtsuki1993,Avishai1993,Gade1993,Ludwig1994,Miller1996,Furusaki1999}.
Therefore, a naive expectation is that $\bm{\phi}_{\frac{N_{\rm u.c.}}{2}}^{\rm NZM}$
exhibits the same scaling behavior as $\bm{u}_{\frac{N_{\rm u.c.}}{2}}$.

In order to examine the scaling behavior of the wave functions in more detail, 
we compute the system size dependence of the IPR.
For the non-zero modes, we define the IPR as 
\begin{eqnarray}
\mathrm{IPR}^{\rm NZM}_{\ell} = \sum_{i} \left[ \rho^{\rm NZM}_{\ell} (i)\right]^2,
\end{eqnarray}
and for the degenerate zero modes, we define it as 
\begin{eqnarray}
\mathrm{IPR}^{\rm ZM} =\sum_{i} \left[ \rho^{\rm ZM} (i)\right]^2.
\end{eqnarray}
The results are shown in Fig.~\ref{fig:ch_ipr}.
Note that we focus on the zero energy modes and the band center, as it is rather clear from Fig.~\ref{fig:ch_probdens}(b) 
that the band edge is a localized state.
We see that the IPR for the zero energy modes is approximately fitted as 
$\mathrm{IPR}^{\rm ZM} \propto L^{-2}$, which is the scaling of the extended state. 
In fact, this is the same behavior as the zero modes of the real-valued random MO model~\cite{Hatsugai2021,Kuroda2022},
and thus is speculated to be a ubiquitous in the zero modes of the random MO models. 
In contrast, for the band center, we see that the IPR is fitted as $\mathrm{IPR}^{\rm NZM}_{\ell = \frac{N_{\rm u.c.}}{2}} \propto L^{-1.76}$,
which is an intermediate behavior between the localized and the extended states. 
Hence, this result supports that the band center is a critical state. 
\begin{figure*}[tb]
\begin{center}
\includegraphics[clip,width = 0.9\linewidth]{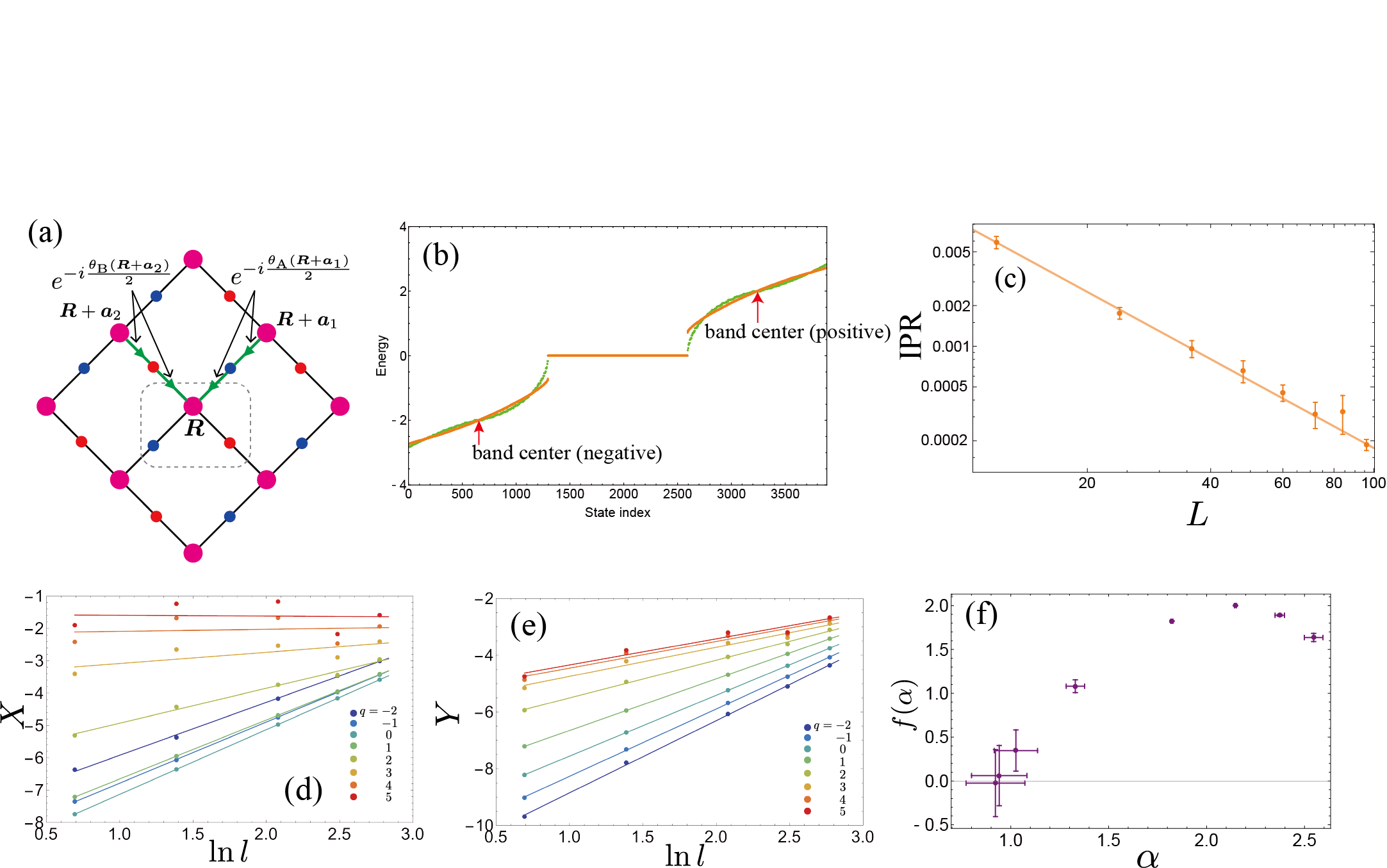}
\vspace{-10pt}
\caption{ 
(a) Schematic figure of the Lieb-lattice model of $\bar{\mathcal{H}}$.
(b) The energy spectrum for the system with $L = 36$. 
Orange (green) dots correspond to the random (clean) system.
(c) IPR  for the center of the positive-energy sector.
The average and standard deviation are calculated for ten samples of the configurations of $\theta$'s. 
The line represent the fitting function, IPR$=A\cdot L^{-B}$,
with $A=0.36$, $B=1.65$. 
(d-f) Multifractal analysis for the center of the positive-energy sector.
(d) $X$ and (e) $Y$ as functions of $\ln l$ for several values of $q$.
(f) $f$-$\alpha$ spectrum. 
The error bars in the panel (f) are the fitting errors 
for $X$ and $Y$.}
  \label{fig:Lieb}
 \end{center}
 \vspace{-10pt}
\end{figure*}

To further shed light on the critical nature of the band center, we perform the multifractal analysis. 
Specifically, we depict the $f$-$\alpha$ spectrum, by which we capture the multifractal nature of the wave function~\cite{Halsey1986}.

Let us briefly summarize how to obtain the $f$-$\alpha$ spectrum, which is based on Refs.~\onlinecite{Chhabra1989_PRL,Chhabra1989}.
First, we cover the system by the patches containing $l \times l$ unit cells. 
We label each patch by $m$.
We then define the probability density in the patch $m$
as 
\begin{eqnarray}
P(l; m) = \sum_{i \in m} \rho^{\rm NZM} (i).
\end{eqnarray}
Note that we omit the mode index $\ell$ for simplicity of writing.
Then, using this, we define 
\begin{eqnarray}
\mu_m (q,l) = \frac{[P (l; m)]^q}{\sum_{m^\prime} [P(l; m^\prime)]^q}.
\end{eqnarray}
The key relations 
to extract the $f$-$\alpha$ spectrum from $\mu_m (q,l)$ are as follows:
\begin{eqnarray}
X (q,l) := \sum_{m} \mu_m(q,l) \ln \mu_m(q,l) = f(q) \ln l,  \label{eq:X}
\end{eqnarray}
and 
\begin{eqnarray}
Y (q,l) := \sum_{m} \mu_m(q,l) \ln P_m(q,l) = \alpha(q) \ln l. \label{eq:Y}
\end{eqnarray}
Using (\ref{eq:X}) and (\ref{eq:Y}), we find that the $f$-$\alpha$ spectrum can be obtained by
(i) plotting $X (q,l) $ and $Y (q,l) $ as functions of $\ln l$ for several values of $q$ 
and (ii) estimating $f(q)$ and $\alpha(q)$ from the slope of the plots of (i).

In Figs.~\ref{fig:ch_fa}(a) and \ref{fig:ch_fa}(b), 
we plot $X$ and $Y$ as functions of $\ln l$, respectively. 
We set $L=96$, $l = 2,4,8,12,16$ and $q = -2,-1,0,1,2,3,4,5$. 
We see that the fitting by using Eqs.~(\ref{eq:X}) and (\ref{eq:Y}) works well for $q\leq 3$,
whereas the deviations of linear fitting for $X$ become relatively large for $q = 4,5 $.
On the basis of this fitting, we obtain the $f$-$\alpha$ spectrum, as shown in Fig.~\ref{fig:ch_fa}(c).
We clearly see the characteristic downward convex behavior of the $f$-$\alpha$ spectrum.
This indicates the multifractal nature of the wave function, which is characteristic to the critical state.

Combining the results of the IPR and that of the multifractal analysis, 
we conclude that the checkerboard model $\mathcal{H}$ hosts the critical state at the center of the finite-energy sector.
Before proceeding further, we remark on a role of symmetry.
In the square-lattice random-phase model, 
it has been argued that the chiral symmetry represented by $\Gamma$ 
(see Sec.~\ref{sec:dual_relation}) plays an essential role in realizing the critical state.
Meanwhile the Hamiltonian $\mathcal{H}$ itself does not preserve the chiral symmetry, 
although the critical state appearing in this model is inherited from the chiral symmetric model, $\tilde{\mathcal{H}}^{\rm sq}$.
In other words, $\mathcal{H}$ serves as an example of the random tight-binding model in unitary class realizing the critical state
without preserving chiral symmetry.

\section{The composite model constructed from molecular orbitals: A Lieb-lattice model \label{sec:lieb}}
\begin{figure}[tb]
\begin{center}
\includegraphics[clip,width = 0.95\linewidth]{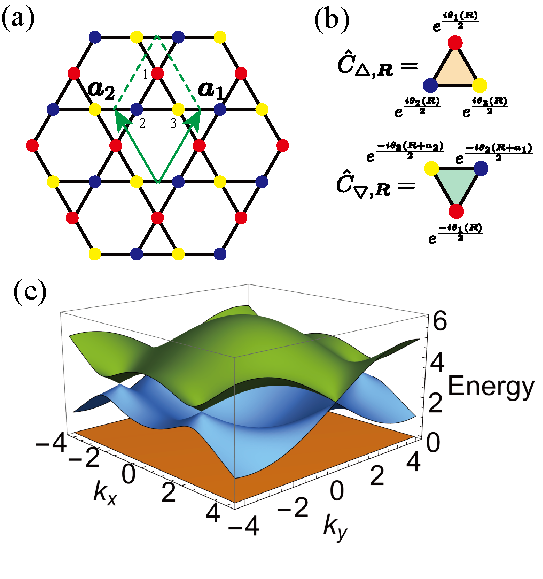}
\vspace{-10pt}
\caption{
(a) Kagome lattice. Two lattice vectors are $\bm{a}_1 = \left(\frac{1}{2},\frac{\sqrt{3}}{2} \right)$ and $\bm{a}_2 = \left(-\frac{1}{2},\frac{\sqrt{3}}{2} \right)$. (b) Schematic figure of $\hat{C}_{\bigtriangleup,\bm{R}} $ and $\hat{C}_{\bigtriangledown,\bm{R}} $.
(c) The dispersion relation for the clean limit [i.e., $\theta_{\rm 1/2/3}(\bm{R}) = 0$].}
  \label{fig:lattice_kagome}
 \end{center}
 \vspace{-10pt}
\end{figure}
\begin{figure*}[tb]
\begin{center}
\includegraphics[clip,width = 0.95\linewidth]{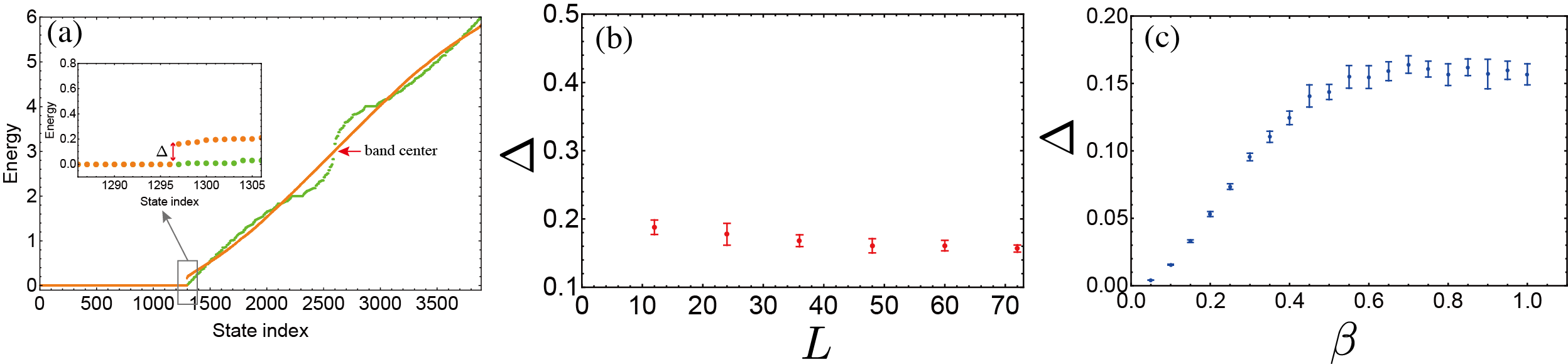}
\vspace{-10pt}
\caption{
(a) Energy spectrum for the system of the kagome lattice model with $L = 36$. 
Orange (green) dots correspond to the random (clean) system.
(b) The gap $\Delta$ as a function of $L$. 
(c) The gap as a function of $\beta$ for $L=60$. 
The average and the standard deviation are calculated for ten independent configurations of $\theta$'s.}
  \label{fig:kagome_en}
 \end{center}
 \vspace{-10pt}
\end{figure*}

So far, we have investigated the checkerboard model constructed by the MOs with random U(1) variable whose Hamiltonian 
matrix is given as $\mathcal{H}$.
There, the existence of the critical state at the center of the finite-energy modess is inherited from that of the 
square-lattice random phase model. 
In this section, we point out that the same trick to obtain the critical state is applicable to 
the composite of the square and the checkerboard lattices, that is, the Lieb lattice. 
In fact, we have already introduced the Lieb-lattice model, $\bar{\mathcal{H}}$, in Eq.~(\ref{eq:Ham_Lieb}) 
to account for the semi-positivity of $\mathcal{H}$. The schematic figure of the Lieb-lattice model is shown in Fig.~\ref{fig:Lieb}(a).

We first remark that the matrix 
$\bar{\mathcal{H}}$ is chiral symmetric, namely, $\bar{\mathcal{H}}$ satisfies
$\{ \bar{\mathcal{H}}, \bar{\Gamma} \} = 0$ with 
\begin{eqnarray}
\bar{\Gamma}  = 
\begin{pmatrix}
\hat{I}_{N_{\rm u.c.}} & \mathcal{O}_{N_{\rm u.c.},2 N_{\rm u.c.}}\\
 \mathcal{O}_{2N_{\rm u.c.},N_{\rm u.c.}}&- \hat{I}_{2N_{\rm u.c.}}  \\
 \end{pmatrix}. \label{eq:chiral_lieb}
\end{eqnarray}
From Eq.~(\ref{eq:chiral_lieb}), we find $| \mathrm{Tr}( \bar{\Gamma})| =N_{\rm u.c.}$.
Due to this relation, there exist $N_{\rm u.c.}$ zero modes~\cite{Lieb1989,Wen1989,Brouwer2002,Koshino2014}.
Also, the chiral symmetry itself implies that the finite-energy states appear at positive and negative in a pairwise manner.
Such behaviors are indeed seen in Fig.~\ref{fig:Lieb}(b), where we plot the energy spectrum of the Lieb-lattice model for $L=36$.
\begin{figure*}[tb]
\begin{center}
\includegraphics[clip,width = 0.95\linewidth]{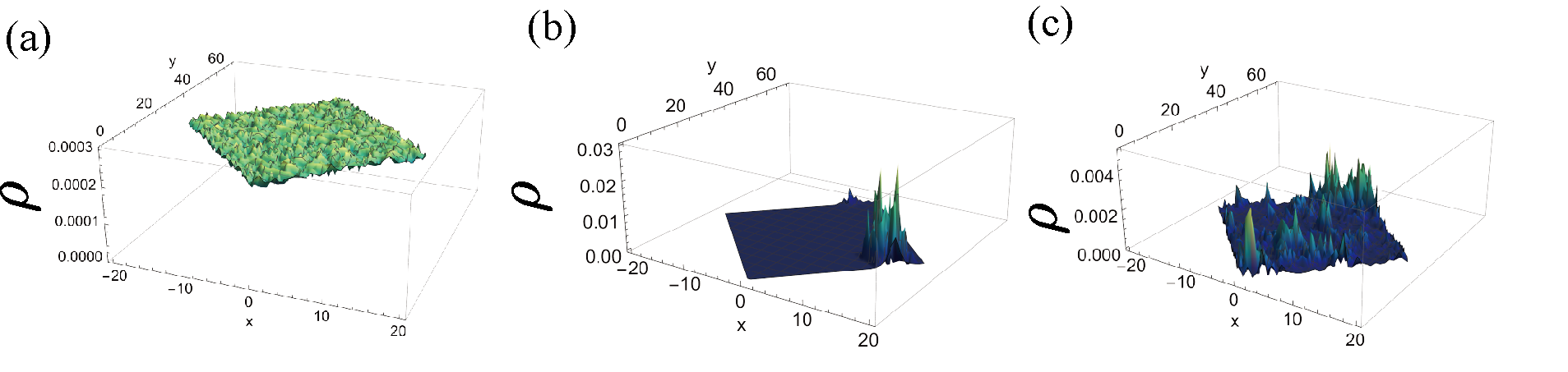}
\vspace{-10pt}
\caption{The probability density for (a) the zero modes, (b) the lowest finite-energy mode,
and (c) the center of the finite-energy modes for the kagome lattice model.
We set $L=36$.}
  \label{fig:k_probdens}
 \end{center}
 \vspace{-10pt}
\end{figure*}

Let us now argue the characters of the wave functions.
Importantly, the finite-energy eigenvectors of $\bar{\mathcal{H}}$ are also constructed from 
the eigenvectors of $\Upsilon$, $\bm{u}_\ell$. 
More precisely, the vector, 
$\bm{\varphi}^{+}_\ell= \frac{1}{\mathcal{N}_{\ell}}\left(\sqrt{\varepsilon_\ell} \bm{u}_\ell, \Psi \bm{u}_\ell \right)^{\rm T}$, 
is the normalized eigenvector of $\bar{\mathcal{H}}$ with the eigenvector $\sqrt{\varepsilon_\ell}$,
where $\mathcal{N}_{\ell}$ is the normalization constant; 
in fact, noting that $|\bm{u}_\ell|^2 = 1$ and $| \Psi \bm{u}_\ell |^2 = \varepsilon_\ell$, one finds $\mathcal{N}_{\ell} = \sqrt{2\varepsilon_\ell}$.
One can easily check this by explicitly multiplying $\bar{\mathcal{H}}$ to $\bm{\varphi}^{+}_\ell$:
\begin{eqnarray}
\bar{\mathcal{H}} \bm{\varphi}^{+}_\ell = \frac{1}{\mathcal{N}_{\ell}}
\begin{pmatrix}
\Psi^\dagger  \Psi \bm{u}_\ell  \\
\sqrt{\varepsilon_\ell}\Psi \bm{u}_\ell \\
\end{pmatrix}
= \frac{\sqrt{\varepsilon_\ell}}{\mathcal{N}_{\ell}} 
\begin{pmatrix}
 \sqrt{\varepsilon_\ell} \bm{u}_\ell  \\
\Psi \bm{u}_\ell \\
\end{pmatrix}
 =  \sqrt{\varepsilon_\ell} \bm{\varphi}^{+}_\ell. \nonumber \\
\end{eqnarray}
Its negative-energy counterpart, i.e., the one whose eigenenergy is $-\sqrt{\varepsilon_\ell}$, 
is given as $\bm{\varphi}^{-}_\ell=\bar{\Gamma} \bm{\varphi}^{-}_\ell= \frac{1}{\mathcal{N}_{\ell}}\left(\sqrt{\varepsilon_\ell} \bm{u}_\ell,- \Psi \bm{u}_\ell \right)^{\rm T}$.
Obviously, $\bm{\varphi}^{-}_\ell$ 
has the same probability density distribution as that of $\bm{\varphi}^{+}_\ell$. 

Then, considering the fact that $\bm{u}_\ell$ and $\Psi \bm{u}_\ell$ are the critical states for $\ell$ being the band center,
we expect that the corresponding $\bm{\varphi}^{+/-}_\ell$ are critical as well. 
We examine the above conjecture by the numerical calculation. 
In Fig.~\ref{fig:Lieb}(c), we plot the IPR for $\bm{\varphi}^{+}_\ell$ for $\ell$ being the band center as a function of $L$.
It exhibits the system size dependence, IPR $\propto L^{-1.65}$, which indicates that the wave function is neither 
localized nor extended.  
Further, in Figs.~\ref{fig:Lieb}(d)-(f), we show the results of the multifractal analysis. 
We see that the deviation from the linear fitting for $X$ for $q = 4,5$ becomes larger than that for the checkerboard model.
In fact, for these values of $q$, $X$ becomes almost independent of the patch size $\ell$.
Nevertheless, the characteristic downward convex behavior of the $f$-$\alpha$ curve is still observed, which 
indicates the multifractal nature of the wave function. 
From these results, we conclude that the Lieb-lattice model $\bar{\mathcal{H}}$ also hosts the critical state.
Additionally, as the probability density of $\bm{\varphi}^{-}_\ell$ is the same as the corresponding $\bm{\varphi}^{+}_\ell$,
the critical state also exists in the negative-energy sector. 
It is noteworthy that the chiral symmetry of $\bar{\mathcal{H}}$ 
is not a direct origin of the emergence of the critical states, because the critical states have finite energy rather than the zero energy. 

\section{Kagome lattice model and  decorated honeycomb lattice model \label{sec:kagome}}
\begin{figure}[tb]
\begin{center}
\includegraphics[clip,width = 0.95\linewidth]{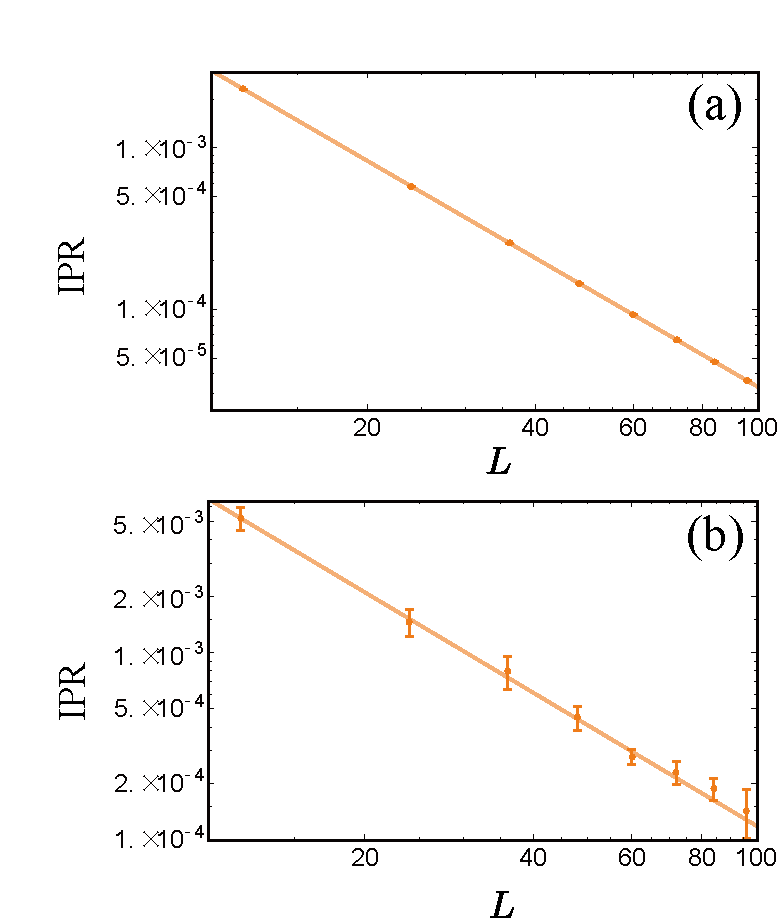}
\vspace{-10pt}
\caption{
IPR for (a) the degenerate zero modes 
and (b) the center of the finite-energy modess
of the kagome lattice model. 
The average and standard deviation is calculated for 10 samples of the configurations of $\theta$'s. 
The line represent the fitting function, IPR$=A\cdot L^{-B}$,
with (a) $A=0.33$, $B=2.00$
and (b) $A=0.43$, $B=1.78$.}
  \label{fig:k_ipr}
 \end{center}
 \vspace{-10pt}
\end{figure}
In this section, we show that the same construction
of the random-phase MO model, discussed in Sec.~\ref{sec:checker}, 
can be applied to the kagome lattice [Fig.~\ref{fig:lattice_kagome}(a)]. 

We consider the tight-binding model:
\begin{eqnarray}
H = \sum_{\bm{R}} \hat{C}^\dagger_{\bigtriangleup, \bm{R}}  \hat{C}_{\bigtriangleup,\bm{R}}  
+  \hat{C}^\dagger_{\bigtriangledown, \bm{R}}  \hat{C}_{\bigtriangledown,\bm{R}} 
\end{eqnarray}
where 
\begin{eqnarray}
 \hat{C}_{\bigtriangleup, \bm{R}} = e^{\frac{i \theta_{\rm 1}(\bm{R})}{2}} c_{\bm{R},\mathrm{1}} 
 + e^{\frac{i \theta_{\rm 2}(\bm{R})}{2}} c_{\bm{R},\mathrm{2}} 
 + e^{\frac{i \theta_{\rm 3}(\bm{R})}{2}} c_{\bm{R},\mathrm{3}},
\end{eqnarray}
and 
\begin{eqnarray}
 \hat{C}_{\bigtriangledown, \bm{R}} = e^{\frac{-i \theta_{\rm 1}(\bm{R})}{2}} c_{\bm{R},\mathrm{1}} 
 + e^{\frac{-i \theta_{\rm 2}(\bm{R}+ \bm{a}_1)}{2}} c_{\bm{R},\mathrm{2}} 
 + e^{\frac{-i \theta_{\rm 3}(\bm{R}+ \bm{a}_2)}{2}} c_{\bm{R},\mathrm{3}}. \nonumber \\
\end{eqnarray}
The phase factors $\theta_{1,2,3}(\bm{R}) \in [-\pi, \pi]$ again obey the uniform distribution. 
See Fig.~\ref{fig:lattice_kagome}(b) for the schematics of $\hat{C}_{\bigtriangleup/\bigtriangledown, \bm{R}}$.
Note that the dual counter part of this model, $\Upsilon$, 
corresponds to the random-phase model on a honeycomb lattice with uniform on-site potential being $3$. 
In the clean limit, we have three bands; the flat band has the lowest energy, 
and the remaining two bands form Dirac cones 
at the $K$ and $K^\prime$ points [Fig.~\ref{fig:lattice_kagome}(c)].

In the following, we show the numerical data for this model.
As we will see, all results are qualitatively the same as that for the checkerboard model.
\begin{figure*}[tb]
\begin{center}
\includegraphics[clip,width = 0.95\linewidth]{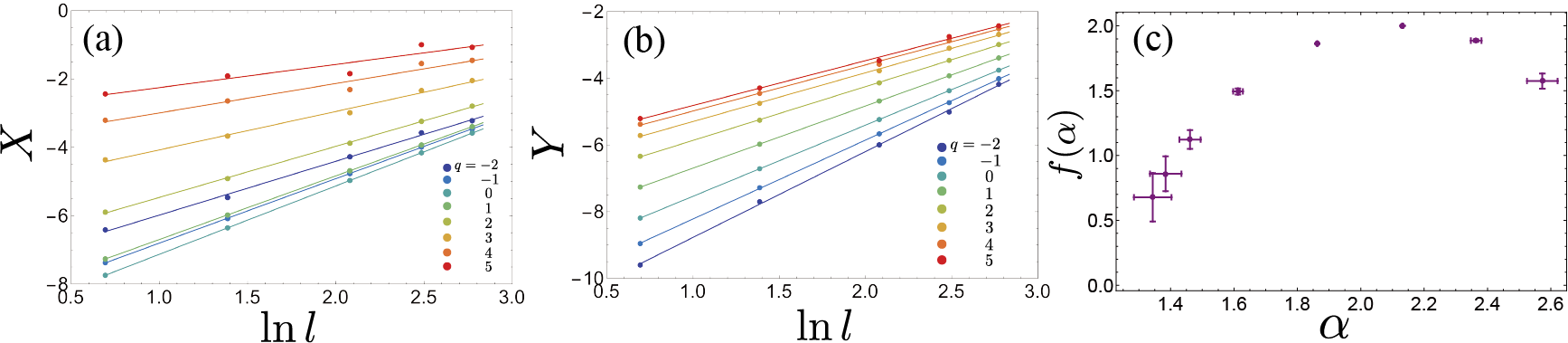}
\vspace{-10pt}
\caption{
Multifractal analysis for the center of the finite-energy modes of the kagome lattice model.
(a) $X$ and (b) $Y$ as functions of $\ln l$ for several values of $q$.
(c) $f$-$\alpha$ spectrum.
The error bars in the panel (c) are the fitting errors 
for $X$ and $Y$.
}
  \label{fig:k_fa}
 \end{center}
 \vspace{-10pt}
\end{figure*}

Figure~\ref{fig:kagome_en}(a) shows the energy spectrum for $L=36$. 
We again see a macroscopically degenerate zero modes even in the presence of the randomness. 
Similarly to the checkerboard model, 
we see a finite gap between the zero modes and the bottom of the finite-energy modes.  
As for the size dependence of the gap [Fig.~\ref{fig:kagome_en}(b)], we again see that the gap 
is expected to be non-vanishing for $L\rightarrow \infty$. 
The gap size is about $0.15$, which is smaller than that for the checkerboard model. 
Additionally, the $\beta$ dependence of the gap [Fig.~\ref{fig:kagome_en}(c)] shows the similar behavior as that for the checkerboard model. 

We next analyze the probability density distribution in the real space. 
Figures~\ref{fig:k_probdens}(a), \ref{fig:k_probdens}(b), and \ref{fig:k_probdens}(c) 
show the probability density distributions
for the zero modes, the band edge, and the band center, respectively. 
We again see qualitatively similar behaviors to the checkerboard model.
In particular, the probability density for the band center seems to have a spiky distribution, which implies the multifractal nature. 
\begin{figure*}[tb]
\begin{center}
\includegraphics[clip,width = 0.9\linewidth]{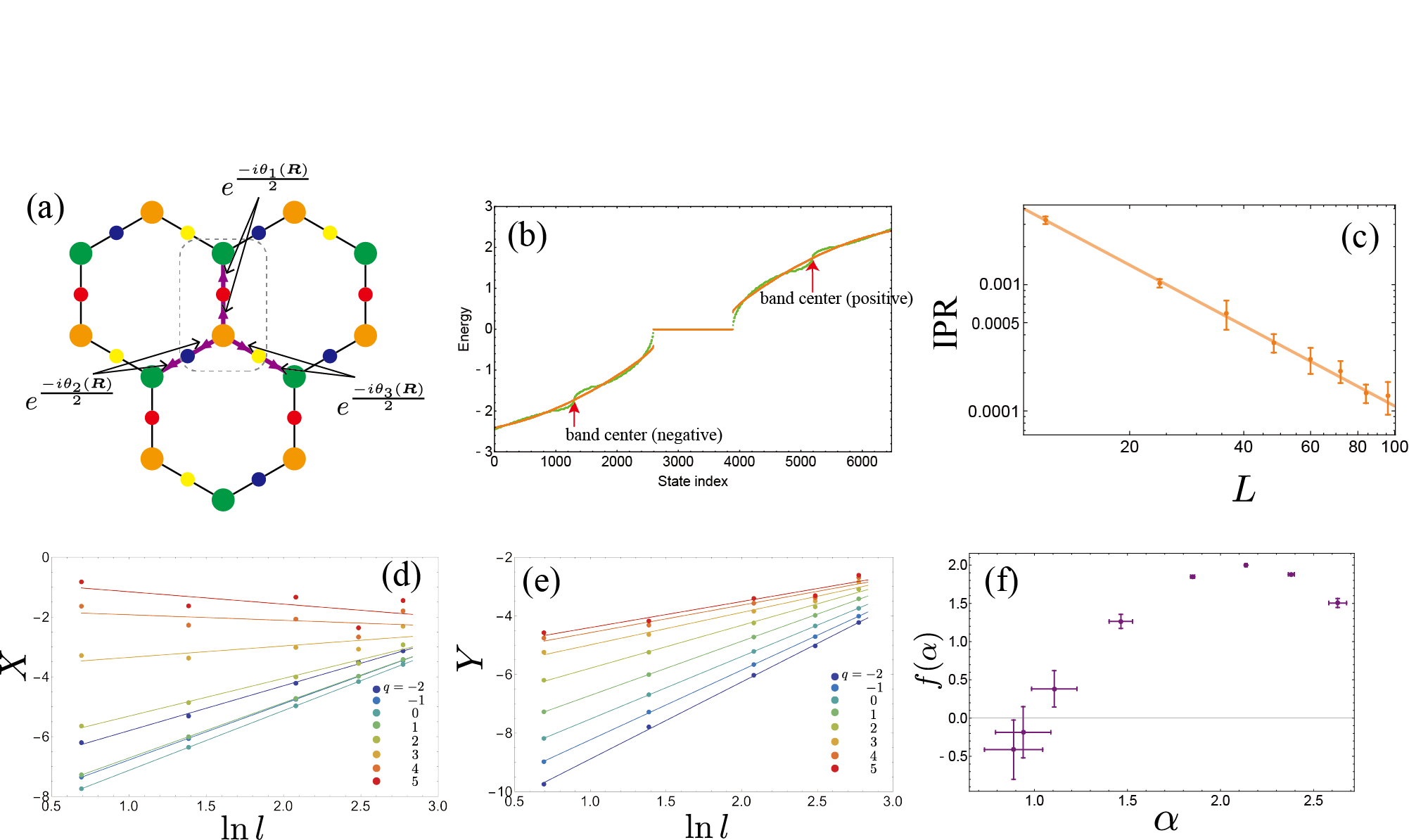}
\vspace{-10pt}
\caption{ 
(a) Schematic figure of the decorated honeycomb lattice model.
(b) The energy spectrum for the system with $L = 36$. 
Orange (green) dots correspond to the random (clean) system.
(c) IPR  for the center of the positive-energy sector.
The average and standard deviation is calculated for 10 samples of the configurations of $\theta$'s. 
The line represent the fitting function, IPR$=A\cdot L^{-B}$,
with $A=0.17$, $B=1.60$.
(d-f) Multifractal analysis for the center of the positive-energy sector.
(d) $X$ and (e) $Y$ as functions of $\ln l$ for several values of $q$.
(f) $f$-$\alpha$ spectrum.
The error bars in the panel (f) are the fitting errors 
for $X$ and $Y$.
}
\label{fig:DH}
\end{center}
\vspace{-10pt}
\end{figure*}

Figures~\ref{fig:k_ipr}(a) and \ref{fig:k_ipr}(b) 
show the system size dependence of the IPR for the zero modes and the band center,
respectively.
We see that the IRP has a scaling behavior of the extended state for the zero modes, 
whereas that for the band center
is of neither the extended state nor the localized state (IPR $\propto L^{-1.78}$). 

Figure~\ref{fig:k_fa} shows the results of the multifractal analysis for a band center. 
We again see that $X$ and $Y$ are linear in $\ln l$, and that the $f$-$\alpha$ curve is downward convex,
indicating the multifractal nature. 

Finally, we introduce a model corresponding to $\bar{\mathcal{H}}$, i.e., the composite of the honeycomb and kagome lattices.
That model is called the decorated honeycomb lattice model, whose schematic figure is shown in Fig.~\ref{fig:DH}(a).
In the energy spectrum [Fig.~\ref{fig:DH}(b)], we see degenerate zero modes and the finite energy modes appearing in a pairwise manner in positive and negative energy sectors. 

Further, as is the case in the Lieb lattice, the band center of the finite-energy modes is again expected to be critical.
This is numerically indicated by the scaling behavior of the IPR [Fig.~\ref{fig:DH}(c)].
As for the results of the multifractal analysis [Figs.~\ref{fig:DH}(d)-(f)], 
we see the same tendency as those of the Lieb lattice [Figs.~\ref{fig:Lieb}(d)-(f)]. 
Specifically, the results for $q = 4,5$ contain large fitting errors, but we believe it reasonable to
judge that the downward convex of the $f$-$\alpha$ spectrum is observed from the data of  $q = -2, \cdots ,3$. 

\section{Summary \label{sec:summary}}
We have proposed a class of tight-binding models constructed by the representation with random U(1) variables.
The model construction scheme guarantees the existence of the macroscopically degenerate zero modes.
A key insight is that the models constructed in this way have their dual counterparts, which are, the random-phase hopping models on bipartite lattices. 
As such, various important properties of eigenvalues and wave functions are closely tied with a dual counterpart. 

By the numerical calculations, we show a gap opening on top of the zero energy modes, which is in contrast to the real-value random MO model. 
The probability density distribution of the zero energy modes 
has a scaling property of the extended state, similar to that of the real-value random MO model.
Besides the zero modes, we also find that the band center of the finite energy sector is the critical state, 
as demonstrated by the scaling of IPR and the characteristic $f$-$\alpha$ spectrum. 
Importantly, the critical state is inherited from its dual counterpart, which is chiral symmetric, 
but the random-phase MO models themselves are lattice models hosting the critical state without preserving the chiral symmetry. 

Furthermore, as a by-product of this model construction scheme, 
we introduce yet another random U(1)-valued hopping model defined on 
a composite lattice such as Lieb and decorated honeycomb lattices,
which host the critical states appearing in a pairwise manner in positive and negative energy sectors.  

Before closing this paper, 
we address several future problems which we believe deserve being studied. 
Firstly, the models we have studied are two-dimensional models, 
but the random U(1) MO models in higher dimensions can be constructed straightforwardly.
Such higher-dimensional models will be a promising target for investigating characteristic localization phenomena, as is the case of the conventional random-phase model~\cite{Ohtsuki1994,Kawarabayashi1998}.
Secondly, the MO representation is applicable to other symmetry classes.
For instance, if we adopt random SU(2) matrices as a coefficients of MOs, 
we obtain random flat-band models in the symplectic class.
Studying localization phenomena in such models, 
for both degenerate zero modes and finite energy modes,
will be an intriguing future problem. 
Finally, experimental realization 
of the random U(1) MO models and the corresponding composite-lattice models is an important issue. 
In fact, the models considered 
in this paper contain only the on-site potentials and the nearest-neighbor hoppings, 
which makes these models feasible in some artificial systems. In cold atoms~\cite{Dalibard2011}, photonic waveguides~\cite{Mukherjee2018}, and electric circuits~\cite{Hofmann2019}, 
the complex hoppings can be implemented, 
so we expect that appropriate control of their amplitudes and phases
will provide chances of realizing our models. 
For instance, in electric circuits, 
systematic control of the admittance of the circuit element 
can be achieved by using the variable resistors and capacitors~\cite{Yatsugi2022},
which will open the way to realize the desired random complex hoppings.  

\acknowledgements
This work is supported by JST CREST, Grant No.~JPMJCR19T1, Japan,
and was partly supported by JSPS KAKENHI, Grants No.~JP17H06138 (T. M. and Y. H.) 
and No.~JP20K14371 (T. M.).

\appendix
\section{Remark on Eq.~(\ref{eq:MO_choice}) \label{app:remark}}
At first glance, the choice of MOs in Eq.~(\ref{eq:MO_choice}) is fine-tuned,
since the phase factors appearing in neighboring MOs are not independent.
An alternative and more naive way of introducing the phase is
\begin{eqnarray}
\hat{\tilde{C}}_{\bm{R}} &=& e^{i \tilde{\theta}_1(\bm{R})} c_{\bm{R},\mathrm{A}} 
 + e^{i \tilde{\theta}_2(\bm{R})} c_{\bm{R},\mathrm{B}}  \nonumber \\
 &+& e^{i \tilde{\theta}_3(\bm{R})} c_{\bm{R}
 +\bm{a}_1,\mathrm{A}} + e^{i \tilde{\theta}_4(\bm{R})} 
 c_{\bm{R}+\bm{a}_2,\mathrm{B}}, 
 \label{eq:MO_choice_2}
\end{eqnarray}
where $\tilde{\theta}_{1,2,3,4}(\bm{R})$ are random variables.
Note that the numbers of $\theta$ variables are $2N_{\rm u.c.}$ for (\ref{eq:MO_choice}) and $4N_{\rm u.c.}$ for (\ref{eq:MO_choice_2}). 
Here we show that (\ref{eq:MO_choice}) and (\ref{eq:MO_choice_2}) gives the same Hamiltonian 
under the gauge transformation. 
Specifically, defining $d_{\bm{R},\mathrm{A}} := e^{\frac{i[\tilde{\theta}_1(\bm{R}) +\tilde{\theta}_3(\bm{R}-\bm{a}_1)]}{2}} c_{\bm{R},\mathrm{A}}$ and $d_{\bm{R},\mathrm{B}} := e^{\frac{i[\tilde{\theta}_2(\bm{R}) +\tilde{\theta}_4(\bm{R}-\bm{a}_2)]}{2}} c_{\bm{R},\mathrm{B}}$,
we have
\begin{widetext}
\begin{eqnarray}
\hat{\tilde{C}}_{\bm{R}} = e^{\frac{i [\tilde{\theta}_1(\bm{R})-\tilde{\theta}_3(\bm{R}-\bm{a}_1)]}{2}} d_{\bm{R},\mathrm{A}} 
 + e^{\frac{i [\tilde{\theta}_2(\bm{R})-\tilde{\theta}_4(\bm{R}-\bm{a}_2)]}{2}}  d_{\bm{R},\mathrm{B}}
 + e^{\frac{-i [\tilde{\theta}_1(\bm{R} + \bm{a}_1)-\tilde{\theta}_3(\bm{R})]}{2}} d_{\bm{R}+\bm{a}_1,\mathrm{A}} 
 + e^{\frac{-i [\tilde{\theta}_2(\bm{R} + \bm{a}_2)-\tilde{\theta}_4(\bm{R})]}{2}} d_{\bm{R}+\bm{a}_2,\mathrm{B}}. 
 \label{eq:MO_choice_3}
\end{eqnarray}
\end{widetext}
Then, by setting $\tilde{\theta}_1(\bm{R})-\tilde{\theta}_3(\bm{R}-\bm{a}_1) = \theta_{\rm A}(\bm{R})$ 
and $\tilde{\theta}_2(\bm{R})-\tilde{\theta}_4(\bm{R}-\bm{a}_2) = \theta_{\rm B}(\bm{R})$, 
we find that $\hat{\tilde{C}}_{\bm{R}}$ of Eq.~(\ref{eq:MO_choice_3}) is equivalent to $\hat{C}_{\bm{R}}$ of Eq.~(\ref{eq:MO_choice}). 

\section{Probability density of finite energy modes \label{app:pd_cm}}
In this appendix, we show that the probability density of the finite energy mode in Eq.~(\ref{eq:PD_NZM})
can be written by using $\Psi$ and the correlation matrix defied for $\bm{u}_\ell$.
This is archived by simply substituting Eq.~(\ref{eq:eigen_h}) into Eq.~(\ref{eq:PD_NZM}):
\begin{eqnarray}
\rho^{\rm NZM}_{\ell} (i) &=& [\bm{\phi}^{\rm NZM}_{\ell} ]_i  [ (\bm{\phi}^{\rm NZM}_{\ell})^\dagger ]_i \nonumber \\
 &=& \frac{1}{\varepsilon_\ell} [\Psi \bm{u}_{\ell} ]_i [ (\bm{u}_{\ell})^\dagger \Psi^\dagger ]_i  \nonumber \\
 &=& \frac{1}{\varepsilon_\ell} [\Psi g^\ell \Psi^\dagger ]_{ii}.
\end{eqnarray}
Here, $g^\ell$ is the $N_{\rm u.c.} \times N_{\rm u.c.} $ matrix,
\begin{eqnarray}
[g^\ell]_{\bm{R},\bm{R}^\prime} = [\bm{u}_\ell ]_{\bm{R}} [\bm{u}^\ast_\ell ]_{\bm{R}^\prime},
\end{eqnarray}
which is referred to as the correlation matrix. 

\bibliographystyle{apsrev4-2}
\bibliography{MO_random_U1}

\begin{thebibliography}{64}%
\makeatletter
\providecommand \@ifxundefined [1]{%
 \@ifx{#1\undefined}
}%
\providecommand \@ifnum [1]{%
 \ifnum #1\expandafter \@firstoftwo
 \else \expandafter \@secondoftwo
 \fi
}%
\providecommand \@ifx [1]{%
 \ifx #1\expandafter \@firstoftwo
 \else \expandafter \@secondoftwo
 \fi
}%
\providecommand \natexlab [1]{#1}%
\providecommand \enquote  [1]{``#1''}%
\providecommand \bibnamefont  [1]{#1}%
\providecommand \bibfnamefont [1]{#1}%
\providecommand \citenamefont [1]{#1}%
\providecommand \href@noop [0]{\@secondoftwo}%
\providecommand \href [0]{\begingroup \@sanitize@url \@href}%
\providecommand \@href[1]{\@@startlink{#1}\@@href}%
\providecommand \@@href[1]{\endgroup#1\@@endlink}%
\providecommand \@sanitize@url [0]{\catcode `\\12\catcode `\$12\catcode
  `\&12\catcode `\#12\catcode `\^12\catcode `\_12\catcode `\%12\relax}%
\providecommand \@@startlink[1]{}%
\providecommand \@@endlink[0]{}%
\providecommand \url  [0]{\begingroup\@sanitize@url \@url }%
\providecommand \@url [1]{\endgroup\@href {#1}{\urlprefix }}%
\providecommand \urlprefix  [0]{URL }%
\providecommand \Eprint [0]{\href }%
\providecommand \doibase [0]{https://doi.org/}%
\providecommand \selectlanguage [0]{\@gobble}%
\providecommand \bibinfo  [0]{\@secondoftwo}%
\providecommand \bibfield  [0]{\@secondoftwo}%
\providecommand \translation [1]{[#1]}%
\providecommand \BibitemOpen [0]{}%
\providecommand \bibitemStop [0]{}%
\providecommand \bibitemNoStop [0]{.\EOS\space}%
\providecommand \EOS [0]{\spacefactor3000\relax}%
\providecommand \BibitemShut  [1]{\csname bibitem#1\endcsname}%
\let\auto@bib@innerbib\@empty
\bibitem [{\citenamefont {Anderson}(1958)}]{Anderson1958}%
  \BibitemOpen
  \bibfield  {author} {\bibinfo {author} {\bibfnamefont {P.~W.}\ \bibnamefont
  {Anderson}},\ }\href {https://doi.org/10.1103/PhysRev.109.1492} {\bibfield
  {journal} {\bibinfo  {journal} {Phys. Rev.}\ }\textbf {\bibinfo {volume}
  {109}},\ \bibinfo {pages} {1492} (\bibinfo {year} {1958})}\BibitemShut
  {NoStop}%
\bibitem [{\citenamefont {Abrahams}\ \emph {et~al.}(1979)\citenamefont
  {Abrahams}, \citenamefont {Anderson}, \citenamefont {Licciardello},\ and\
  \citenamefont {Ramakrishnan}}]{Abrahams1979}%
  \BibitemOpen
  \bibfield  {author} {\bibinfo {author} {\bibfnamefont {E.}~\bibnamefont
  {Abrahams}}, \bibinfo {author} {\bibfnamefont {P.~W.}\ \bibnamefont
  {Anderson}}, \bibinfo {author} {\bibfnamefont {D.~C.}\ \bibnamefont
  {Licciardello}},\ and\ \bibinfo {author} {\bibfnamefont {T.~V.}\ \bibnamefont
  {Ramakrishnan}},\ }\href {https://doi.org/10.1103/PhysRevLett.42.673}
  {\bibfield  {journal} {\bibinfo  {journal} {Phys. Rev. Lett.}\ }\textbf
  {\bibinfo {volume} {42}},\ \bibinfo {pages} {673} (\bibinfo {year}
  {1979})}\BibitemShut {NoStop}%
\bibitem [{\citenamefont {Hatsugai}\ and\ \citenamefont
  {Lee}(1993)}]{Hatsugai1993}%
  \BibitemOpen
  \bibfield  {author} {\bibinfo {author} {\bibfnamefont {Y.}~\bibnamefont
  {Hatsugai}}\ and\ \bibinfo {author} {\bibfnamefont {P.~A.}\ \bibnamefont
  {Lee}},\ }\href {https://doi.org/10.1103/PhysRevB.48.4204} {\bibfield
  {journal} {\bibinfo  {journal} {Phys. Rev. B}\ }\textbf {\bibinfo {volume}
  {48}},\ \bibinfo {pages} {4204} (\bibinfo {year} {1993})}\BibitemShut
  {NoStop}%
\bibitem [{\citenamefont {Chamon}\ \emph {et~al.}(1996)\citenamefont {Chamon},
  \citenamefont {Mudry},\ and\ \citenamefont {Wen}}]{Chamon1996}%
  \BibitemOpen
  \bibfield  {author} {\bibinfo {author} {\bibfnamefont {C.~d.~C.}\
  \bibnamefont {Chamon}}, \bibinfo {author} {\bibfnamefont {C.}~\bibnamefont
  {Mudry}},\ and\ \bibinfo {author} {\bibfnamefont {X.-G.}\ \bibnamefont
  {Wen}},\ }\href {https://doi.org/10.1103/PhysRevB.53.R7638} {\bibfield
  {journal} {\bibinfo  {journal} {Phys. Rev. B}\ }\textbf {\bibinfo {volume}
  {53}},\ \bibinfo {pages} {R7638} (\bibinfo {year} {1996})}\BibitemShut
  {NoStop}%
\bibitem [{\citenamefont {Hatsugai}\ \emph {et~al.}(1997)\citenamefont
  {Hatsugai}, \citenamefont {Wen},\ and\ \citenamefont
  {Kohmoto}}]{Hatsugai1997}%
  \BibitemOpen
  \bibfield  {author} {\bibinfo {author} {\bibfnamefont {Y.}~\bibnamefont
  {Hatsugai}}, \bibinfo {author} {\bibfnamefont {X.-G.}\ \bibnamefont {Wen}},\
  and\ \bibinfo {author} {\bibfnamefont {M.}~\bibnamefont {Kohmoto}},\ }\href
  {https://doi.org/10.1103/PhysRevB.56.1061} {\bibfield  {journal} {\bibinfo
  {journal} {Phys. Rev. B}\ }\textbf {\bibinfo {volume} {56}},\ \bibinfo
  {pages} {1061} (\bibinfo {year} {1997})}\BibitemShut {NoStop}%
\bibitem [{\citenamefont {Castillo}\ \emph {et~al.}(1997)\citenamefont
  {Castillo}, \citenamefont {de~C.~Chamon}, \citenamefont {Fradkin},
  \citenamefont {Goldbart},\ and\ \citenamefont {Mudry}}]{Castillo1997}%
  \BibitemOpen
  \bibfield  {author} {\bibinfo {author} {\bibfnamefont {H.~E.}\ \bibnamefont
  {Castillo}}, \bibinfo {author} {\bibfnamefont {C.}~\bibnamefont
  {de~C.~Chamon}}, \bibinfo {author} {\bibfnamefont {E.}~\bibnamefont
  {Fradkin}}, \bibinfo {author} {\bibfnamefont {P.~M.}\ \bibnamefont
  {Goldbart}},\ and\ \bibinfo {author} {\bibfnamefont {C.}~\bibnamefont
  {Mudry}},\ }\href {https://doi.org/10.1103/PhysRevB.56.10668} {\bibfield
  {journal} {\bibinfo  {journal} {Phys. Rev. B}\ }\textbf {\bibinfo {volume}
  {56}},\ \bibinfo {pages} {10668} (\bibinfo {year} {1997})}\BibitemShut
  {NoStop}%
\bibitem [{\citenamefont {Morita}\ and\ \citenamefont
  {Hatsugai}(1997)}]{Morita1997}%
  \BibitemOpen
  \bibfield  {author} {\bibinfo {author} {\bibfnamefont {Y.}~\bibnamefont
  {Morita}}\ and\ \bibinfo {author} {\bibfnamefont {Y.}~\bibnamefont
  {Hatsugai}},\ }\href {https://doi.org/10.1103/PhysRevLett.79.3728} {\bibfield
   {journal} {\bibinfo  {journal} {Phys. Rev. Lett.}\ }\textbf {\bibinfo
  {volume} {79}},\ \bibinfo {pages} {3728} (\bibinfo {year}
  {1997})}\BibitemShut {NoStop}%
\bibitem [{\citenamefont {Fukui}(2003)}]{Fukui2003}%
  \BibitemOpen
  \bibfield  {author} {\bibinfo {author} {\bibfnamefont {T.}~\bibnamefont
  {Fukui}},\ }\href {https://doi.org/10.1103/PhysRevB.68.153307} {\bibfield
  {journal} {\bibinfo  {journal} {Phys. Rev. B}\ }\textbf {\bibinfo {volume}
  {68}},\ \bibinfo {pages} {153307} (\bibinfo {year} {2003})}\BibitemShut
  {NoStop}%
\bibitem [{\citenamefont {Zhitomirsky}\ and\ \citenamefont
  {Tsunetsugu}(2004)}]{Zhitomirsky2004}%
  \BibitemOpen
  \bibfield  {author} {\bibinfo {author} {\bibfnamefont {M.~E.}\ \bibnamefont
  {Zhitomirsky}}\ and\ \bibinfo {author} {\bibfnamefont {H.}~\bibnamefont
  {Tsunetsugu}},\ }\href {https://doi.org/10.1103/PhysRevB.70.100403}
  {\bibfield  {journal} {\bibinfo  {journal} {Phys. Rev. B}\ }\textbf {\bibinfo
  {volume} {70}},\ \bibinfo {pages} {100403} (\bibinfo {year}
  {2004})}\BibitemShut {NoStop}%
\bibitem [{\citenamefont {Bergman}\ \emph {et~al.}(2008)\citenamefont
  {Bergman}, \citenamefont {Wu},\ and\ \citenamefont {Balents}}]{Bergman2008}%
  \BibitemOpen
  \bibfield  {author} {\bibinfo {author} {\bibfnamefont {D.~L.}\ \bibnamefont
  {Bergman}}, \bibinfo {author} {\bibfnamefont {C.}~\bibnamefont {Wu}},\ and\
  \bibinfo {author} {\bibfnamefont {L.}~\bibnamefont {Balents}},\ }\href
  {https://doi.org/10.1103/PhysRevB.78.125104} {\bibfield  {journal} {\bibinfo
  {journal} {Phys. Rev. B}\ }\textbf {\bibinfo {volume} {78}},\ \bibinfo
  {pages} {125104} (\bibinfo {year} {2008})}\BibitemShut {NoStop}%
\bibitem [{\citenamefont {Kuno}\ \emph {et~al.}(2020)\citenamefont {Kuno},
  \citenamefont {Mizoguchi},\ and\ \citenamefont {Hatsugai}}]{Kuno2020}%
  \BibitemOpen
  \bibfield  {author} {\bibinfo {author} {\bibfnamefont {Y.}~\bibnamefont
  {Kuno}}, \bibinfo {author} {\bibfnamefont {T.}~\bibnamefont {Mizoguchi}},\
  and\ \bibinfo {author} {\bibfnamefont {Y.}~\bibnamefont {Hatsugai}},\ }\href
  {https://doi.org/10.1103/PhysRevB.102.241115} {\bibfield  {journal} {\bibinfo
   {journal} {Phys. Rev. B}\ }\textbf {\bibinfo {volume} {102}},\ \bibinfo
  {pages} {241115} (\bibinfo {year} {2020})}\BibitemShut {NoStop}%
\bibitem [{\citenamefont {Goda}\ \emph {et~al.}(2006)\citenamefont {Goda},
  \citenamefont {Nishino},\ and\ \citenamefont {Matsuda}}]{Goda2006}%
  \BibitemOpen
  \bibfield  {author} {\bibinfo {author} {\bibfnamefont {M.}~\bibnamefont
  {Goda}}, \bibinfo {author} {\bibfnamefont {S.}~\bibnamefont {Nishino}},\ and\
  \bibinfo {author} {\bibfnamefont {H.}~\bibnamefont {Matsuda}},\ }\href
  {https://doi.org/10.1103/PhysRevLett.96.126401} {\bibfield  {journal}
  {\bibinfo  {journal} {Phys. Rev. Lett.}\ }\textbf {\bibinfo {volume} {96}},\
  \bibinfo {pages} {126401} (\bibinfo {year} {2006})}\BibitemShut {NoStop}%
\bibitem [{\citenamefont {Nishino}\ \emph {et~al.}(2007)\citenamefont
  {Nishino}, \citenamefont {Matsuda},\ and\ \citenamefont
  {Goda}}]{Nishino2007}%
  \BibitemOpen
  \bibfield  {author} {\bibinfo {author} {\bibfnamefont {S.}~\bibnamefont
  {Nishino}}, \bibinfo {author} {\bibfnamefont {H.}~\bibnamefont {Matsuda}},\
  and\ \bibinfo {author} {\bibfnamefont {M.}~\bibnamefont {Goda}},\ }\href
  {https://doi.org/10.1143/JPSJ.76.024709} {\bibfield  {journal} {\bibinfo
  {journal} {Journal of the Physical Society of Japan}\ }\textbf {\bibinfo
  {volume} {76}},\ \bibinfo {pages} {024709} (\bibinfo {year}
  {2007})}\BibitemShut {NoStop}%
\bibitem [{\citenamefont {Chalker}\ \emph {et~al.}(2010)\citenamefont
  {Chalker}, \citenamefont {Pickles},\ and\ \citenamefont
  {Shukla}}]{Chalker2010}%
  \BibitemOpen
  \bibfield  {author} {\bibinfo {author} {\bibfnamefont {J.~T.}\ \bibnamefont
  {Chalker}}, \bibinfo {author} {\bibfnamefont {T.~S.}\ \bibnamefont
  {Pickles}},\ and\ \bibinfo {author} {\bibfnamefont {P.}~\bibnamefont
  {Shukla}},\ }\href {https://doi.org/10.1103/PhysRevB.82.104209} {\bibfield
  {journal} {\bibinfo  {journal} {Phys. Rev. B}\ }\textbf {\bibinfo {volume}
  {82}},\ \bibinfo {pages} {104209} (\bibinfo {year} {2010})}\BibitemShut
  {NoStop}%
\bibitem [{\citenamefont {Leykam}\ \emph {et~al.}(2013)\citenamefont {Leykam},
  \citenamefont {Flach}, \citenamefont {Bahat-Treidel},\ and\ \citenamefont
  {Desyatnikov}}]{Leykam2013}%
  \BibitemOpen
  \bibfield  {author} {\bibinfo {author} {\bibfnamefont {D.}~\bibnamefont
  {Leykam}}, \bibinfo {author} {\bibfnamefont {S.}~\bibnamefont {Flach}},
  \bibinfo {author} {\bibfnamefont {O.}~\bibnamefont {Bahat-Treidel}},\ and\
  \bibinfo {author} {\bibfnamefont {A.~S.}\ \bibnamefont {Desyatnikov}},\
  }\href {https://doi.org/10.1103/PhysRevB.88.224203} {\bibfield  {journal}
  {\bibinfo  {journal} {Phys. Rev. B}\ }\textbf {\bibinfo {volume} {88}},\
  \bibinfo {pages} {224203} (\bibinfo {year} {2013})}\BibitemShut {NoStop}%
\bibitem [{\citenamefont {Shukla}(2018{\natexlab{a}})}]{Shukla2018}%
  \BibitemOpen
  \bibfield  {author} {\bibinfo {author} {\bibfnamefont {P.}~\bibnamefont
  {Shukla}},\ }\href {https://doi.org/10.1103/PhysRevB.98.054206} {\bibfield
  {journal} {\bibinfo  {journal} {Phys. Rev. B}\ }\textbf {\bibinfo {volume}
  {98}},\ \bibinfo {pages} {054206} (\bibinfo {year}
  {2018}{\natexlab{a}})}\BibitemShut {NoStop}%
\bibitem [{\citenamefont {Shukla}(2018{\natexlab{b}})}]{Shukla2018_2}%
  \BibitemOpen
  \bibfield  {author} {\bibinfo {author} {\bibfnamefont {P.}~\bibnamefont
  {Shukla}},\ }\href {https://doi.org/10.1103/PhysRevB.98.184202} {\bibfield
  {journal} {\bibinfo  {journal} {Phys. Rev. B}\ }\textbf {\bibinfo {volume}
  {98}},\ \bibinfo {pages} {184202} (\bibinfo {year}
  {2018}{\natexlab{b}})}\BibitemShut {NoStop}%
\bibitem [{\citenamefont {Bilitewski}\ and\ \citenamefont
  {Moessner}(2018)}]{Bilitewski2018}%
  \BibitemOpen
  \bibfield  {author} {\bibinfo {author} {\bibfnamefont {T.}~\bibnamefont
  {Bilitewski}}\ and\ \bibinfo {author} {\bibfnamefont {R.}~\bibnamefont
  {Moessner}},\ }\href {https://doi.org/10.1103/PhysRevB.98.235109} {\bibfield
  {journal} {\bibinfo  {journal} {Phys. Rev. B}\ }\textbf {\bibinfo {volume}
  {98}},\ \bibinfo {pages} {235109} (\bibinfo {year} {2018})}\BibitemShut
  {NoStop}%
\bibitem [{\citenamefont {Kuno}\ \emph {et~al.}(2021)\citenamefont {Kuno},
  \citenamefont {Mizoguchi},\ and\ \citenamefont {Hatsugai}}]{Kuno2021}%
  \BibitemOpen
  \bibfield  {author} {\bibinfo {author} {\bibfnamefont {Y.}~\bibnamefont
  {Kuno}}, \bibinfo {author} {\bibfnamefont {T.}~\bibnamefont {Mizoguchi}},\
  and\ \bibinfo {author} {\bibfnamefont {Y.}~\bibnamefont {Hatsugai}},\ }\href
  {https://doi.org/10.1103/PhysRevB.104.085130} {\bibfield  {journal} {\bibinfo
   {journal} {Phys. Rev. B}\ }\textbf {\bibinfo {volume} {104}},\ \bibinfo
  {pages} {085130} (\bibinfo {year} {2021})}\BibitemShut {NoStop}%
\bibitem [{\citenamefont {Altland}\ and\ \citenamefont
  {Zirnbauer}(1997)}]{Altland1997}%
  \BibitemOpen
  \bibfield  {author} {\bibinfo {author} {\bibfnamefont {A.}~\bibnamefont
  {Altland}}\ and\ \bibinfo {author} {\bibfnamefont {M.~R.}\ \bibnamefont
  {Zirnbauer}},\ }\href {https://doi.org/10.1103/PhysRevB.55.1142} {\bibfield
  {journal} {\bibinfo  {journal} {Phys. Rev. B}\ }\textbf {\bibinfo {volume}
  {55}},\ \bibinfo {pages} {1142} (\bibinfo {year} {1997})}\BibitemShut
  {NoStop}%
\bibitem [{rem({\natexlab{a}})}]{remark}%
  \BibitemOpen
  \href@noop {} {}\bibinfo {note} {For simplicity, we assume that there is one
  AO at each lattice site. Extension to spinful and/or multiorbital cases will
  be an intriguing future problem.}\BibitemShut {Stop}%
\bibitem [{\citenamefont {Hatsugai}\ and\ \citenamefont
  {Maruyama}(2011)}]{Hatsugai2011}%
  \BibitemOpen
  \bibfield  {author} {\bibinfo {author} {\bibfnamefont {Y.}~\bibnamefont
  {Hatsugai}}\ and\ \bibinfo {author} {\bibfnamefont {I.}~\bibnamefont
  {Maruyama}},\ }\href {https://doi.org/10.1209/0295-5075/95/20003} {\bibfield
  {journal} {\bibinfo  {journal} {{EPL} (Europhysics Letters)}\ }\textbf
  {\bibinfo {volume} {95}},\ \bibinfo {pages} {20003} (\bibinfo {year}
  {2011})}\BibitemShut {NoStop}%
\bibitem [{\citenamefont {Hatsugai}\ \emph {et~al.}(2015)\citenamefont
  {Hatsugai}, \citenamefont {Shiraishi},\ and\ \citenamefont
  {Aoki}}]{Hatsugai2015}%
  \BibitemOpen
  \bibfield  {author} {\bibinfo {author} {\bibfnamefont {Y.}~\bibnamefont
  {Hatsugai}}, \bibinfo {author} {\bibfnamefont {K.}~\bibnamefont
  {Shiraishi}},\ and\ \bibinfo {author} {\bibfnamefont {H.}~\bibnamefont
  {Aoki}},\ }\href {https://doi.org/10.1088/1367-2630/17/2/025009} {\bibfield
  {journal} {\bibinfo  {journal} {New Journal of Physics}\ }\textbf {\bibinfo
  {volume} {17}},\ \bibinfo {pages} {025009} (\bibinfo {year}
  {2015})}\BibitemShut {NoStop}%
\bibitem [{\citenamefont {Mizoguchi}\ and\ \citenamefont
  {Hatsugai}(2019)}]{Mizoguchi2019}%
  \BibitemOpen
  \bibfield  {author} {\bibinfo {author} {\bibfnamefont {T.}~\bibnamefont
  {Mizoguchi}}\ and\ \bibinfo {author} {\bibfnamefont {Y.}~\bibnamefont
  {Hatsugai}},\ }\href {https://doi.org/10.1209/0295-5075/127/47001} {\bibfield
   {journal} {\bibinfo  {journal} {{EPL} (Europhysics Letters)}\ }\textbf
  {\bibinfo {volume} {127}},\ \bibinfo {pages} {47001} (\bibinfo {year}
  {2019})}\BibitemShut {NoStop}%
\bibitem [{\citenamefont {Mizoguchi}\ and\ \citenamefont
  {Hatsugai}(2020)}]{Mizoguchi2020}%
  \BibitemOpen
  \bibfield  {author} {\bibinfo {author} {\bibfnamefont {T.}~\bibnamefont
  {Mizoguchi}}\ and\ \bibinfo {author} {\bibfnamefont {Y.}~\bibnamefont
  {Hatsugai}},\ }\href {https://doi.org/10.1103/PhysRevB.101.235125} {\bibfield
   {journal} {\bibinfo  {journal} {Phys. Rev. B}\ }\textbf {\bibinfo {volume}
  {101}},\ \bibinfo {pages} {235125} (\bibinfo {year} {2020})}\BibitemShut
  {NoStop}%
\bibitem [{\citenamefont {Mizoguchi}\ \emph
  {et~al.}(2021{\natexlab{a}})\citenamefont {Mizoguchi}, \citenamefont {Kuno},\
  and\ \citenamefont {Hatsugai}}]{Mizoguchi2021_skagome}%
  \BibitemOpen
  \bibfield  {author} {\bibinfo {author} {\bibfnamefont {T.}~\bibnamefont
  {Mizoguchi}}, \bibinfo {author} {\bibfnamefont {Y.}~\bibnamefont {Kuno}},\
  and\ \bibinfo {author} {\bibfnamefont {Y.}~\bibnamefont {Hatsugai}},\ }\href
  {https://doi.org/10.1103/PhysRevB.104.035161} {\bibfield  {journal} {\bibinfo
   {journal} {Phys. Rev. B}\ }\textbf {\bibinfo {volume} {104}},\ \bibinfo
  {pages} {035161} (\bibinfo {year} {2021}{\natexlab{a}})}\BibitemShut
  {NoStop}%
\bibitem [{\citenamefont {Hatsugai}(2021)}]{Hatsugai2021}%
  \BibitemOpen
  \bibfield  {author} {\bibinfo {author} {\bibfnamefont {Y.}~\bibnamefont
  {Hatsugai}},\ }\href
  {https://doi.org/https://doi.org/10.1016/j.aop.2021.168453} {\bibfield
  {journal} {\bibinfo  {journal} {Annals of Physics}\ }\textbf {\bibinfo
  {volume} {435}},\ \bibinfo {pages} {168453} (\bibinfo {year} {2021})},\
  \bibinfo {note} {special Issue on Localisation 2020}\BibitemShut {NoStop}%
\bibitem [{\citenamefont {Mizoguchi}\ \emph {et~al.}(2022)\citenamefont
  {Mizoguchi}, \citenamefont {Kuno},\ and\ \citenamefont
  {Hatsugai}}]{Mizoguchi2022}%
  \BibitemOpen
  \bibfield  {author} {\bibinfo {author} {\bibfnamefont {T.}~\bibnamefont
  {Mizoguchi}}, \bibinfo {author} {\bibfnamefont {Y.}~\bibnamefont {Kuno}},\
  and\ \bibinfo {author} {\bibfnamefont {Y.}~\bibnamefont {Hatsugai}},\
  }\bibfield  {journal} {\bibinfo  {journal} {Progress of Theoretical and
  Experimental Physics}\ }\textbf {\bibinfo {volume} {2022}},\ \href
  {https://doi.org/10.1093/ptep/ptac015} {10.1093/ptep/ptac015} (\bibinfo
  {year} {2022}),\ \bibinfo {note} {023I02}\BibitemShut {NoStop}%
\bibitem [{\citenamefont {Kuroda}\ \emph {et~al.}(2022)\citenamefont {Kuroda},
  \citenamefont {Mizoguchi}, \citenamefont {Araki},\ and\ \citenamefont
  {Hatsugai}}]{Kuroda2022}%
  \BibitemOpen
  \bibfield  {author} {\bibinfo {author} {\bibfnamefont {T.}~\bibnamefont
  {Kuroda}}, \bibinfo {author} {\bibfnamefont {T.}~\bibnamefont {Mizoguchi}},
  \bibinfo {author} {\bibfnamefont {H.}~\bibnamefont {Araki}},\ and\ \bibinfo
  {author} {\bibfnamefont {Y.}~\bibnamefont {Hatsugai}},\ }\href
  {https://doi.org/10.7566/JPSJ.91.044703} {\bibfield  {journal} {\bibinfo
  {journal} {Journal of the Physical Society of Japan}\ }\textbf {\bibinfo
  {volume} {91}},\ \bibinfo {pages} {044703} (\bibinfo {year}
  {2022})}\BibitemShut {NoStop}%
\bibitem [{rem({\natexlab{b}})}]{remark_mo}%
  \BibitemOpen
  \href@noop {} {}\bibinfo {note} {More precisely, in the models we have
  studied so far, the amplitudes of the real-valued coefficients are
  random.}\BibitemShut {Stop}%
\bibitem [{\citenamefont {Lee}\ and\ \citenamefont {Fisher}(1981)}]{Lee1981}%
  \BibitemOpen
  \bibfield  {author} {\bibinfo {author} {\bibfnamefont {P.~A.}\ \bibnamefont
  {Lee}}\ and\ \bibinfo {author} {\bibfnamefont {D.~S.}\ \bibnamefont
  {Fisher}},\ }\href {https://doi.org/10.1103/PhysRevLett.47.882} {\bibfield
  {journal} {\bibinfo  {journal} {Phys. Rev. Lett.}\ }\textbf {\bibinfo
  {volume} {47}},\ \bibinfo {pages} {882} (\bibinfo {year} {1981})}\BibitemShut
  {NoStop}%
\bibitem [{\citenamefont {Pryor}\ and\ \citenamefont {Zee}(1992)}]{Pryor1992}%
  \BibitemOpen
  \bibfield  {author} {\bibinfo {author} {\bibfnamefont {C.}~\bibnamefont
  {Pryor}}\ and\ \bibinfo {author} {\bibfnamefont {A.}~\bibnamefont {Zee}},\
  }\href {https://doi.org/10.1103/PhysRevB.46.3116} {\bibfield  {journal}
  {\bibinfo  {journal} {Phys. Rev. B}\ }\textbf {\bibinfo {volume} {46}},\
  \bibinfo {pages} {3116} (\bibinfo {year} {1992})}\BibitemShut {NoStop}%
\bibitem [{\citenamefont {Kalmeyer}\ \emph {et~al.}(1993)\citenamefont
  {Kalmeyer}, \citenamefont {Wei}, \citenamefont {Arovas},\ and\ \citenamefont
  {Zhang}}]{Kalmeyer1993}%
  \BibitemOpen
  \bibfield  {author} {\bibinfo {author} {\bibfnamefont {V.}~\bibnamefont
  {Kalmeyer}}, \bibinfo {author} {\bibfnamefont {D.}~\bibnamefont {Wei}},
  \bibinfo {author} {\bibfnamefont {D.~P.}\ \bibnamefont {Arovas}},\ and\
  \bibinfo {author} {\bibfnamefont {S.}~\bibnamefont {Zhang}},\ }\href
  {https://doi.org/10.1103/PhysRevB.48.11095} {\bibfield  {journal} {\bibinfo
  {journal} {Phys. Rev. B}\ }\textbf {\bibinfo {volume} {48}},\ \bibinfo
  {pages} {11095} (\bibinfo {year} {1993})}\BibitemShut {NoStop}%
\bibitem [{\citenamefont {Sugiyama}\ and\ \citenamefont
  {Nagaosa}(1993)}]{Sugiyama1993}%
  \BibitemOpen
  \bibfield  {author} {\bibinfo {author} {\bibfnamefont {T.}~\bibnamefont
  {Sugiyama}}\ and\ \bibinfo {author} {\bibfnamefont {N.}~\bibnamefont
  {Nagaosa}},\ }\href {https://doi.org/10.1103/PhysRevLett.70.1980} {\bibfield
  {journal} {\bibinfo  {journal} {Phys. Rev. Lett.}\ }\textbf {\bibinfo
  {volume} {70}},\ \bibinfo {pages} {1980} (\bibinfo {year}
  {1993})}\BibitemShut {NoStop}%
\bibitem [{\citenamefont {Ohtsuki}\ \emph {et~al.}(1993)\citenamefont
  {Ohtsuki}, \citenamefont {Slevin},\ and\ \citenamefont {Ono}}]{Ohtsuki1993}%
  \BibitemOpen
  \bibfield  {author} {\bibinfo {author} {\bibfnamefont {T.}~\bibnamefont
  {Ohtsuki}}, \bibinfo {author} {\bibfnamefont {K.}~\bibnamefont {Slevin}},\
  and\ \bibinfo {author} {\bibfnamefont {Y.}~\bibnamefont {Ono}},\ }\href
  {https://doi.org/10.1143/JPSJ.62.3979} {\bibfield  {journal} {\bibinfo
  {journal} {Journal of the Physical Society of Japan}\ }\textbf {\bibinfo
  {volume} {62}},\ \bibinfo {pages} {3979} (\bibinfo {year}
  {1993})}\BibitemShut {NoStop}%
\bibitem [{\citenamefont {Avishai}\ \emph {et~al.}(1993)\citenamefont
  {Avishai}, \citenamefont {Hatsugai},\ and\ \citenamefont
  {Kohmoto}}]{Avishai1993}%
  \BibitemOpen
  \bibfield  {author} {\bibinfo {author} {\bibfnamefont {Y.}~\bibnamefont
  {Avishai}}, \bibinfo {author} {\bibfnamefont {Y.}~\bibnamefont {Hatsugai}},\
  and\ \bibinfo {author} {\bibfnamefont {M.}~\bibnamefont {Kohmoto}},\ }\href
  {https://doi.org/10.1103/PhysRevB.47.9561} {\bibfield  {journal} {\bibinfo
  {journal} {Phys. Rev. B}\ }\textbf {\bibinfo {volume} {47}},\ \bibinfo
  {pages} {9561} (\bibinfo {year} {1993})}\BibitemShut {NoStop}%
\bibitem [{\citenamefont {Gade}(1993)}]{Gade1993}%
  \BibitemOpen
  \bibfield  {author} {\bibinfo {author} {\bibfnamefont {R.}~\bibnamefont
  {Gade}},\ }\href
  {https://doi.org/https://doi.org/10.1016/0550-3213(93)90601-K} {\bibfield
  {journal} {\bibinfo  {journal} {Nuclear Physics B}\ }\textbf {\bibinfo
  {volume} {398}},\ \bibinfo {pages} {499} (\bibinfo {year}
  {1993})}\BibitemShut {NoStop}%
\bibitem [{\citenamefont {Ludwig}\ \emph {et~al.}(1994)\citenamefont {Ludwig},
  \citenamefont {Fisher}, \citenamefont {Shankar},\ and\ \citenamefont
  {Grinstein}}]{Ludwig1994}%
  \BibitemOpen
  \bibfield  {author} {\bibinfo {author} {\bibfnamefont {A.~W.~W.}\
  \bibnamefont {Ludwig}}, \bibinfo {author} {\bibfnamefont {M.~P.~A.}\
  \bibnamefont {Fisher}}, \bibinfo {author} {\bibfnamefont {R.}~\bibnamefont
  {Shankar}},\ and\ \bibinfo {author} {\bibfnamefont {G.}~\bibnamefont
  {Grinstein}},\ }\href {https://doi.org/10.1103/PhysRevB.50.7526} {\bibfield
  {journal} {\bibinfo  {journal} {Phys. Rev. B}\ }\textbf {\bibinfo {volume}
  {50}},\ \bibinfo {pages} {7526} (\bibinfo {year} {1994})}\BibitemShut
  {NoStop}%
\bibitem [{\citenamefont {Miller}\ and\ \citenamefont
  {Wang}(1996)}]{Miller1996}%
  \BibitemOpen
  \bibfield  {author} {\bibinfo {author} {\bibfnamefont {J.}~\bibnamefont
  {Miller}}\ and\ \bibinfo {author} {\bibfnamefont {J.}~\bibnamefont {Wang}},\
  }\href {https://doi.org/10.1103/PhysRevLett.76.1461} {\bibfield  {journal}
  {\bibinfo  {journal} {Phys. Rev. Lett.}\ }\textbf {\bibinfo {volume} {76}},\
  \bibinfo {pages} {1461} (\bibinfo {year} {1996})}\BibitemShut {NoStop}%
\bibitem [{\citenamefont {Furusaki}(1999)}]{Furusaki1999}%
  \BibitemOpen
  \bibfield  {author} {\bibinfo {author} {\bibfnamefont {A.}~\bibnamefont
  {Furusaki}},\ }\href {https://doi.org/10.1103/PhysRevLett.82.604} {\bibfield
  {journal} {\bibinfo  {journal} {Phys. Rev. Lett.}\ }\textbf {\bibinfo
  {volume} {82}},\ \bibinfo {pages} {604} (\bibinfo {year} {1999})}\BibitemShut
  {NoStop}%
\bibitem [{\citenamefont {Mizoguchi}\ \emph {et~al.}(2020)\citenamefont
  {Mizoguchi}, \citenamefont {Kuno},\ and\ \citenamefont
  {Hatsugai}}]{Mizoguchi2020_sq}%
  \BibitemOpen
  \bibfield  {author} {\bibinfo {author} {\bibfnamefont {T.}~\bibnamefont
  {Mizoguchi}}, \bibinfo {author} {\bibfnamefont {Y.}~\bibnamefont {Kuno}},\
  and\ \bibinfo {author} {\bibfnamefont {Y.}~\bibnamefont {Hatsugai}},\ }\href
  {https://doi.org/10.1103/PhysRevA.102.033527} {\bibfield  {journal} {\bibinfo
   {journal} {Phys. Rev. A}\ }\textbf {\bibinfo {volume} {102}},\ \bibinfo
  {pages} {033527} (\bibinfo {year} {2020})}\BibitemShut {NoStop}%
\bibitem [{\citenamefont {Mizoguchi}\ \emph
  {et~al.}(2021{\natexlab{b}})\citenamefont {Mizoguchi}, \citenamefont {Kuno},\
  and\ \citenamefont {Hatsugai}}]{Mizoguchi_erratum}%
  \BibitemOpen
  \bibfield  {author} {\bibinfo {author} {\bibfnamefont {T.}~\bibnamefont
  {Mizoguchi}}, \bibinfo {author} {\bibfnamefont {Y.}~\bibnamefont {Kuno}},\
  and\ \bibinfo {author} {\bibfnamefont {Y.}~\bibnamefont {Hatsugai}},\ }\href
  {https://doi.org/10.1103/PhysRevA.104.029906} {\bibfield  {journal} {\bibinfo
   {journal} {Phys. Rev. A}\ }\textbf {\bibinfo {volume} {104}},\ \bibinfo
  {pages} {029906} (\bibinfo {year} {2021}{\natexlab{b}})}\BibitemShut
  {NoStop}%
\bibitem [{\citenamefont {Mizoguchi}\ \emph
  {et~al.}(2021{\natexlab{c}})\citenamefont {Mizoguchi}, \citenamefont
  {Yoshida},\ and\ \citenamefont {Hatsugai}}]{Mizoguchi2021}%
  \BibitemOpen
  \bibfield  {author} {\bibinfo {author} {\bibfnamefont {T.}~\bibnamefont
  {Mizoguchi}}, \bibinfo {author} {\bibfnamefont {T.}~\bibnamefont {Yoshida}},\
  and\ \bibinfo {author} {\bibfnamefont {Y.}~\bibnamefont {Hatsugai}},\ }\href
  {https://doi.org/10.1103/PhysRevB.103.045136} {\bibfield  {journal} {\bibinfo
   {journal} {Phys. Rev. B}\ }\textbf {\bibinfo {volume} {103}},\ \bibinfo
  {pages} {045136} (\bibinfo {year} {2021}{\natexlab{c}})}\BibitemShut
  {NoStop}%
\bibitem [{\citenamefont {Rhim}\ and\ \citenamefont {Yang}(2019)}]{Rhim2019}%
  \BibitemOpen
  \bibfield  {author} {\bibinfo {author} {\bibfnamefont {J.-W.}\ \bibnamefont
  {Rhim}}\ and\ \bibinfo {author} {\bibfnamefont {B.-J.}\ \bibnamefont
  {Yang}},\ }\href {https://doi.org/10.1103/PhysRevB.99.045107} {\bibfield
  {journal} {\bibinfo  {journal} {Phys. Rev. B}\ }\textbf {\bibinfo {volume}
  {99}},\ \bibinfo {pages} {045107} (\bibinfo {year} {2019})}\BibitemShut
  {NoStop}%
\bibitem [{\citenamefont {Hwang}\ \emph
  {et~al.}(2021{\natexlab{a}})\citenamefont {Hwang}, \citenamefont {Rhim},\
  and\ \citenamefont {Yang}}]{Hwang2021}%
  \BibitemOpen
  \bibfield  {author} {\bibinfo {author} {\bibfnamefont {Y.}~\bibnamefont
  {Hwang}}, \bibinfo {author} {\bibfnamefont {J.-W.}\ \bibnamefont {Rhim}},\
  and\ \bibinfo {author} {\bibfnamefont {B.-J.}\ \bibnamefont {Yang}},\ }\href
  {https://doi.org/10.1103/PhysRevB.104.L081104} {\bibfield  {journal}
  {\bibinfo  {journal} {Phys. Rev. B}\ }\textbf {\bibinfo {volume} {104}},\
  \bibinfo {pages} {L081104} (\bibinfo {year}
  {2021}{\natexlab{a}})}\BibitemShut {NoStop}%
\bibitem [{\citenamefont {Hwang}\ \emph
  {et~al.}(2021{\natexlab{b}})\citenamefont {Hwang}, \citenamefont {Rhim},\
  and\ \citenamefont {Yang}}]{Hwang2021_2}%
  \BibitemOpen
  \bibfield  {author} {\bibinfo {author} {\bibfnamefont {Y.}~\bibnamefont
  {Hwang}}, \bibinfo {author} {\bibfnamefont {J.-W.}\ \bibnamefont {Rhim}},\
  and\ \bibinfo {author} {\bibfnamefont {B.-J.}\ \bibnamefont {Yang}},\ }\href
  {https://doi.org/10.1103/PhysRevB.104.085144} {\bibfield  {journal} {\bibinfo
   {journal} {Phys. Rev. B}\ }\textbf {\bibinfo {volume} {104}},\ \bibinfo
  {pages} {085144} (\bibinfo {year} {2021}{\natexlab{b}})}\BibitemShut
  {NoStop}%
\bibitem [{\citenamefont {Graf}\ and\ \citenamefont
  {Pi\'echon}(2021)}]{Graf2021}%
  \BibitemOpen
  \bibfield  {author} {\bibinfo {author} {\bibfnamefont {A.}~\bibnamefont
  {Graf}}\ and\ \bibinfo {author} {\bibfnamefont {F.}~\bibnamefont
  {Pi\'echon}},\ }\href {https://doi.org/10.1103/PhysRevB.104.195128}
  {\bibfield  {journal} {\bibinfo  {journal} {Phys. Rev. B}\ }\textbf {\bibinfo
  {volume} {104}},\ \bibinfo {pages} {195128} (\bibinfo {year}
  {2021})}\BibitemShut {NoStop}%
\bibitem [{\citenamefont {{Di Francesco}}\ and\ \citenamefont
  {Zuber}(1990)}]{DiFrancesco1990}%
  \BibitemOpen
  \bibfield  {author} {\bibinfo {author} {\bibfnamefont {P.}~\bibnamefont {{Di
  Francesco}}}\ and\ \bibinfo {author} {\bibfnamefont {J.-B.}\ \bibnamefont
  {Zuber}},\ }\href
  {https://doi.org/https://doi.org/10.1016/0550-3213(90)90645-T} {\bibfield
  {journal} {\bibinfo  {journal} {Nuclear Physics B}\ }\textbf {\bibinfo
  {volume} {338}},\ \bibinfo {pages} {602} (\bibinfo {year}
  {1990})}\BibitemShut {NoStop}%
\bibitem [{\citenamefont {Pearce}\ and\ \citenamefont
  {Zhou}(1993)}]{Pearce1993}%
  \BibitemOpen
  \bibfield  {author} {\bibinfo {author} {\bibfnamefont {P.~A.}\ \bibnamefont
  {Pearce}}\ and\ \bibinfo {author} {\bibfnamefont {Y.-K.}\ \bibnamefont
  {Zhou}},\ }\href {https://doi.org/10.1142/S0217979293003449} {\bibfield
  {journal} {\bibinfo  {journal} {International Journal of Modern Physics B}\
  }\textbf {\bibinfo {volume} {07}},\ \bibinfo {pages} {3649} (\bibinfo {year}
  {1993})}\BibitemShut {NoStop}%
\bibitem [{\citenamefont {Mizoguchi}\ \emph
  {et~al.}(2021{\natexlab{d}})\citenamefont {Mizoguchi}, \citenamefont
  {Katsura}, \citenamefont {Maruyama},\ and\ \citenamefont
  {Hatsugai}}]{Mizoguchi2021_Diamond}%
  \BibitemOpen
  \bibfield  {author} {\bibinfo {author} {\bibfnamefont {T.}~\bibnamefont
  {Mizoguchi}}, \bibinfo {author} {\bibfnamefont {H.}~\bibnamefont {Katsura}},
  \bibinfo {author} {\bibfnamefont {I.}~\bibnamefont {Maruyama}},\ and\
  \bibinfo {author} {\bibfnamefont {Y.}~\bibnamefont {Hatsugai}},\ }\href
  {https://doi.org/10.1103/PhysRevB.104.035155} {\bibfield  {journal} {\bibinfo
   {journal} {Phys. Rev. B}\ }\textbf {\bibinfo {volume} {104}},\ \bibinfo
  {pages} {035155} (\bibinfo {year} {2021}{\natexlab{d}})}\BibitemShut
  {NoStop}%
\bibitem [{rem({\natexlab{c}})}]{remark_chiral}%
  \BibitemOpen
  \href@noop {} {}\bibinfo {note} {Strictly speaking, under the periodic
  boundary condition, the chiral symmetry is preserved only when $L$ is
  even.}\BibitemShut {Stop}%
\bibitem [{\citenamefont {Halsey}\ \emph {et~al.}(1986)\citenamefont {Halsey},
  \citenamefont {Jensen}, \citenamefont {Kadanoff}, \citenamefont {Procaccia},\
  and\ \citenamefont {Shraiman}}]{Halsey1986}%
  \BibitemOpen
  \bibfield  {author} {\bibinfo {author} {\bibfnamefont {T.~C.}\ \bibnamefont
  {Halsey}}, \bibinfo {author} {\bibfnamefont {M.~H.}\ \bibnamefont {Jensen}},
  \bibinfo {author} {\bibfnamefont {L.~P.}\ \bibnamefont {Kadanoff}}, \bibinfo
  {author} {\bibfnamefont {I.}~\bibnamefont {Procaccia}},\ and\ \bibinfo
  {author} {\bibfnamefont {B.~I.}\ \bibnamefont {Shraiman}},\ }\href
  {https://doi.org/10.1103/PhysRevA.33.1141} {\bibfield  {journal} {\bibinfo
  {journal} {Phys. Rev. A}\ }\textbf {\bibinfo {volume} {33}},\ \bibinfo
  {pages} {1141} (\bibinfo {year} {1986})}\BibitemShut {NoStop}%
\bibitem [{\citenamefont {Chhabra}\ and\ \citenamefont
  {Jensen}(1989)}]{Chhabra1989_PRL}%
  \BibitemOpen
  \bibfield  {author} {\bibinfo {author} {\bibfnamefont {A.}~\bibnamefont
  {Chhabra}}\ and\ \bibinfo {author} {\bibfnamefont {R.~V.}\ \bibnamefont
  {Jensen}},\ }\href {https://doi.org/10.1103/PhysRevLett.62.1327} {\bibfield
  {journal} {\bibinfo  {journal} {Phys. Rev. Lett.}\ }\textbf {\bibinfo
  {volume} {62}},\ \bibinfo {pages} {1327} (\bibinfo {year}
  {1989})}\BibitemShut {NoStop}%
\bibitem [{\citenamefont {Chhabra}\ \emph {et~al.}(1989)\citenamefont
  {Chhabra}, \citenamefont {Meneveau}, \citenamefont {Jensen},\ and\
  \citenamefont {Sreenivasan}}]{Chhabra1989}%
  \BibitemOpen
  \bibfield  {author} {\bibinfo {author} {\bibfnamefont {A.~B.}\ \bibnamefont
  {Chhabra}}, \bibinfo {author} {\bibfnamefont {C.}~\bibnamefont {Meneveau}},
  \bibinfo {author} {\bibfnamefont {R.~V.}\ \bibnamefont {Jensen}},\ and\
  \bibinfo {author} {\bibfnamefont {K.~R.}\ \bibnamefont {Sreenivasan}},\
  }\href {https://doi.org/10.1103/PhysRevA.40.5284} {\bibfield  {journal}
  {\bibinfo  {journal} {Phys. Rev. A}\ }\textbf {\bibinfo {volume} {40}},\
  \bibinfo {pages} {5284} (\bibinfo {year} {1989})}\BibitemShut {NoStop}%
\bibitem [{\citenamefont {Lieb}(1989)}]{Lieb1989}%
  \BibitemOpen
  \bibfield  {author} {\bibinfo {author} {\bibfnamefont {E.~H.}\ \bibnamefont
  {Lieb}},\ }\href {https://doi.org/10.1103/PhysRevLett.62.1201} {\bibfield
  {journal} {\bibinfo  {journal} {Phys. Rev. Lett.}\ }\textbf {\bibinfo
  {volume} {62}},\ \bibinfo {pages} {1201} (\bibinfo {year}
  {1989})}\BibitemShut {NoStop}%
\bibitem [{\citenamefont {Wen}\ and\ \citenamefont {Zee}(1989)}]{Wen1989}%
  \BibitemOpen
  \bibfield  {author} {\bibinfo {author} {\bibfnamefont {X.}~\bibnamefont
  {Wen}}\ and\ \bibinfo {author} {\bibfnamefont {A.}~\bibnamefont {Zee}},\
  }\href {https://doi.org/https://doi.org/10.1016/0550-3213(89)90062-X}
  {\bibfield  {journal} {\bibinfo  {journal} {Nuclear Physics B}\ }\textbf
  {\bibinfo {volume} {316}},\ \bibinfo {pages} {641} (\bibinfo {year}
  {1989})}\BibitemShut {NoStop}%
\bibitem [{\citenamefont {Brouwer}\ \emph {et~al.}(2002)\citenamefont
  {Brouwer}, \citenamefont {Racine}, \citenamefont {Furusaki}, \citenamefont
  {Hatsugai}, \citenamefont {Morita},\ and\ \citenamefont
  {Mudry}}]{Brouwer2002}%
  \BibitemOpen
  \bibfield  {author} {\bibinfo {author} {\bibfnamefont {P.~W.}\ \bibnamefont
  {Brouwer}}, \bibinfo {author} {\bibfnamefont {E.}~\bibnamefont {Racine}},
  \bibinfo {author} {\bibfnamefont {A.}~\bibnamefont {Furusaki}}, \bibinfo
  {author} {\bibfnamefont {Y.}~\bibnamefont {Hatsugai}}, \bibinfo {author}
  {\bibfnamefont {Y.}~\bibnamefont {Morita}},\ and\ \bibinfo {author}
  {\bibfnamefont {C.}~\bibnamefont {Mudry}},\ }\href
  {https://doi.org/10.1103/PhysRevB.66.014204} {\bibfield  {journal} {\bibinfo
  {journal} {Phys. Rev. B}\ }\textbf {\bibinfo {volume} {66}},\ \bibinfo
  {pages} {014204} (\bibinfo {year} {2002})}\BibitemShut {NoStop}%
\bibitem [{\citenamefont {Koshino}\ \emph {et~al.}(2014)\citenamefont
  {Koshino}, \citenamefont {Morimoto},\ and\ \citenamefont
  {Sato}}]{Koshino2014}%
  \BibitemOpen
  \bibfield  {author} {\bibinfo {author} {\bibfnamefont {M.}~\bibnamefont
  {Koshino}}, \bibinfo {author} {\bibfnamefont {T.}~\bibnamefont {Morimoto}},\
  and\ \bibinfo {author} {\bibfnamefont {M.}~\bibnamefont {Sato}},\ }\href
  {https://doi.org/10.1103/PhysRevB.90.115207} {\bibfield  {journal} {\bibinfo
  {journal} {Phys. Rev. B}\ }\textbf {\bibinfo {volume} {90}},\ \bibinfo
  {pages} {115207} (\bibinfo {year} {2014})}\BibitemShut {NoStop}%
\bibitem [{\citenamefont {Ohtsuki}\ \emph {et~al.}(1994)\citenamefont
  {Ohtsuki}, \citenamefont {Ono},\ and\ \citenamefont {Kramer}}]{Ohtsuki1994}%
  \BibitemOpen
  \bibfield  {author} {\bibinfo {author} {\bibfnamefont {T.}~\bibnamefont
  {Ohtsuki}}, \bibinfo {author} {\bibfnamefont {Y.}~\bibnamefont {Ono}},\ and\
  \bibinfo {author} {\bibfnamefont {B.}~\bibnamefont {Kramer}},\ }\href
  {https://doi.org/10.1143/JPSJ.63.685} {\bibfield  {journal} {\bibinfo
  {journal} {Journal of the Physical Society of Japan}\ }\textbf {\bibinfo
  {volume} {63}},\ \bibinfo {pages} {685} (\bibinfo {year} {1994})}\BibitemShut
  {NoStop}%
\bibitem [{\citenamefont {Kawarabayashi}\ \emph {et~al.}(1998)\citenamefont
  {Kawarabayashi}, \citenamefont {Kramer},\ and\ \citenamefont
  {Ohtsuki}}]{Kawarabayashi1998}%
  \BibitemOpen
  \bibfield  {author} {\bibinfo {author} {\bibfnamefont {T.}~\bibnamefont
  {Kawarabayashi}}, \bibinfo {author} {\bibfnamefont {B.}~\bibnamefont
  {Kramer}},\ and\ \bibinfo {author} {\bibfnamefont {T.}~\bibnamefont
  {Ohtsuki}},\ }\href {https://doi.org/10.1103/PhysRevB.57.11842} {\bibfield
  {journal} {\bibinfo  {journal} {Phys. Rev. B}\ }\textbf {\bibinfo {volume}
  {57}},\ \bibinfo {pages} {11842} (\bibinfo {year} {1998})}\BibitemShut
  {NoStop}%
\bibitem [{\citenamefont {Dalibard}\ \emph {et~al.}(2011)\citenamefont
  {Dalibard}, \citenamefont {Gerbier}, \citenamefont
  {Juzeli\ifmmode~\bar{u}\else \={u}\fi{}nas},\ and\ \citenamefont
  {\"Ohberg}}]{Dalibard2011}%
  \BibitemOpen
  \bibfield  {author} {\bibinfo {author} {\bibfnamefont {J.}~\bibnamefont
  {Dalibard}}, \bibinfo {author} {\bibfnamefont {F.}~\bibnamefont {Gerbier}},
  \bibinfo {author} {\bibfnamefont {G.}~\bibnamefont
  {Juzeli\ifmmode~\bar{u}\else \={u}\fi{}nas}},\ and\ \bibinfo {author}
  {\bibfnamefont {P.}~\bibnamefont {\"Ohberg}},\ }\href
  {https://doi.org/10.1103/RevModPhys.83.1523} {\bibfield  {journal} {\bibinfo
  {journal} {Rev. Mod. Phys.}\ }\textbf {\bibinfo {volume} {83}},\ \bibinfo
  {pages} {1523} (\bibinfo {year} {2011})}\BibitemShut {NoStop}%
\bibitem [{\citenamefont {Mukherjee}\ \emph {et~al.}(2018)\citenamefont
  {Mukherjee}, \citenamefont {Di~Liberto}, \citenamefont {\"Ohberg},
  \citenamefont {Thomson},\ and\ \citenamefont {Goldman}}]{Mukherjee2018}%
  \BibitemOpen
  \bibfield  {author} {\bibinfo {author} {\bibfnamefont {S.}~\bibnamefont
  {Mukherjee}}, \bibinfo {author} {\bibfnamefont {M.}~\bibnamefont
  {Di~Liberto}}, \bibinfo {author} {\bibfnamefont {P.}~\bibnamefont
  {\"Ohberg}}, \bibinfo {author} {\bibfnamefont {R.~R.}\ \bibnamefont
  {Thomson}},\ and\ \bibinfo {author} {\bibfnamefont {N.}~\bibnamefont
  {Goldman}},\ }\href {https://doi.org/10.1103/PhysRevLett.121.075502}
  {\bibfield  {journal} {\bibinfo  {journal} {Phys. Rev. Lett.}\ }\textbf
  {\bibinfo {volume} {121}},\ \bibinfo {pages} {075502} (\bibinfo {year}
  {2018})}\BibitemShut {NoStop}%
\bibitem [{\citenamefont {Hofmann}\ \emph {et~al.}(2019)\citenamefont
  {Hofmann}, \citenamefont {Helbig}, \citenamefont {Lee}, \citenamefont
  {Greiter},\ and\ \citenamefont {Thomale}}]{Hofmann2019}%
  \BibitemOpen
  \bibfield  {author} {\bibinfo {author} {\bibfnamefont {T.}~\bibnamefont
  {Hofmann}}, \bibinfo {author} {\bibfnamefont {T.}~\bibnamefont {Helbig}},
  \bibinfo {author} {\bibfnamefont {C.~H.}\ \bibnamefont {Lee}}, \bibinfo
  {author} {\bibfnamefont {M.}~\bibnamefont {Greiter}},\ and\ \bibinfo {author}
  {\bibfnamefont {R.}~\bibnamefont {Thomale}},\ }\href
  {https://doi.org/10.1103/PhysRevLett.122.247702} {\bibfield  {journal}
  {\bibinfo  {journal} {Phys. Rev. Lett.}\ }\textbf {\bibinfo {volume} {122}},\
  \bibinfo {pages} {247702} (\bibinfo {year} {2019})}\BibitemShut {NoStop}%
\bibitem [{\citenamefont {Yatsugi}\ \emph {et~al.}(2022)\citenamefont
  {Yatsugi}, \citenamefont {Yoshida}, \citenamefont {Mizoguchi}, \citenamefont
  {Kuno}, \citenamefont {Iizuka}, \citenamefont {Tadokoro},\ and\ \citenamefont
  {Hatsugai}}]{Yatsugi2022}%
  \BibitemOpen
  \bibfield  {author} {\bibinfo {author} {\bibfnamefont {K.}~\bibnamefont
  {Yatsugi}}, \bibinfo {author} {\bibfnamefont {T.}~\bibnamefont {Yoshida}},
  \bibinfo {author} {\bibfnamefont {T.}~\bibnamefont {Mizoguchi}}, \bibinfo
  {author} {\bibfnamefont {Y.}~\bibnamefont {Kuno}}, \bibinfo {author}
  {\bibfnamefont {H.}~\bibnamefont {Iizuka}}, \bibinfo {author} {\bibfnamefont
  {Y.}~\bibnamefont {Tadokoro}},\ and\ \bibinfo {author} {\bibfnamefont
  {Y.}~\bibnamefont {Hatsugai}},\ }\href
  {https://doi.org/10.1038/s42005-022-00957-5} {\bibfield  {journal} {\bibinfo
  {journal} {Communications Physics}\ }\textbf {\bibinfo {volume} {5}},\
  \bibinfo {pages} {180} (\bibinfo {year} {2022})}\BibitemShut {NoStop}%
\end{thebibliography}%
\end{document}